\documentclass[11pt,a4paper]{article} 
\usepackage{jcappub}

\usepackage{color}

\usepackage{amsmath}
\usepackage{amsfonts}
\usepackage{graphicx}
\usepackage{hyperref}
\usepackage{natbib}
\usepackage{bm}

\usepackage{varioref}
\usepackage{amssymb}
\usepackage{comment}

\newcommand{\scN}{\textsc{n}}

\renewcommand\thesection{\arabic{section}}

\newcommand{\be}{\begin{equation}}
\newcommand{\ee}{\end{equation}}
\newcommand{\bea}{\begin{eqnarray}} 
\newcommand{\eea}{\end{eqnarray}}
\newcommand{\ft}[2]{{\textstyle\frac{#1}{#2}}}

\newsavebox{\uuunit}
\sbox{\uuunit}
    {\setlength{\unitlength}{0.825em}
     \begin{picture}(0.6,0.7)
        \thinlines
        \put(0,0){\line(1,0){0.5}}
        \put(0.15,0){\line(0,1){0.7}}
        \put(0.35,0){\line(0,1){0.8}}
       \multiput(0.3,0.8)(-0.04,-0.02){12}{\rule{0.5pt}{0.5pt}}
     \end {picture}}
\newcommand {\unity}{\mathord{\!\usebox{\uuunit}}}

 \def\cM{{\cal M}}   
  \def\cR{{\cal R}}

 \def\cS{{\cal S}}

\def\atH0{|_{H_{0}}}
\def\AtH0{\bigg|_{H_{0}}}

\csname @addtoreset\endcsname{equation}{section}

\DeclareMathOperator\erf{erf}

\title{Large Scale Power Suppression in a Multifield Landscape}

\author{Jose J. Blanco-Pillado${}^{a,b}$, Mafalda Dias${}^{c}$, Jonathan Frazer${}^{a,b}$, and Kepa Sousa${}^{a}$}
\affiliation{$^a$ Department of Theoretical Physics, University of the Basque Country,  Bilbao, Spain}
\affiliation{$^b$ IKERBASQUE, Basque Foundation for Science, 48011, Bilbao, Spain}
\affiliation{$^c$Astronomy Centre, University of Sussex, Falmer, Brighton, BN1 9QH, UK}

\abstract{Power suppression of the cosmic microwave background on the largest observable scales could provide valuable clues about the particle physics underlying inflation. Here we consider the prospect of power suppression in the context of the multifield landscape. Based on the assumption that our observable universe emerges from a tunnelling event and that the relevant features originate purely from inflationary dynamics, we find that the power spectrum not only contains information on single-field dynamics, but also places strong constraints on all scalar fields present in the theory. We find that the simplest single-field models giving rise to power suppression do not generalise to multifield models in a straightforward way, as the resulting superhorizon evolution of the curvature perturbation tends to erase any power suppression present at horizon crossing. On the other hand, multifield effects do present a means of generating power suppression which to our knowledge has so far not been considered. We propose a mechanism to illustrate this, which we dub \emph{flume inflation.}}

\begin{document}

\maketitle
\flushbottom

\section{Introduction}

There is substantial evidence both on theoretical and observational fronts indicating that our observable universe underwent 
a period of inflation \cite{Guth:1980zm,Sato:1980yn,Linde:1981mu,Albrecht:1982wi}. If inflation really occurred, it provides 
an extraordinary opportunity to use observations of the Cosmic Microwave Background (CMB) and large scale structure
as a means of studying particle physics at energy scales well above those likely to be achieved with terrestrial experiments. 
In particular, inflation is sensitive to Planck suppressed operators, allowing us to test 
ultraviolet complete theories in a unique way.  It remains a central challenge for inflation to understand how it is to be 
embedded in such a theory but there has been significant progress in the context of string theory (for a recent review see \cite{1404.2601}). 

Developments in the understanding of compactification over the last decade have given rise to a striking picture that 
should have profound implications for the study of inflation --- the prevalence of large Hodge numbers 
\cite{hep-th/0610102, hep-th/0701050, 0803.1194, hep-th/0509003} indicates the existence of many, often 
hundreds of scalar fields, interacting via a complicated potential containing a large number of metastable vacua. 
This picture is sometimes referred to as the string landscape \cite{Bousso:2000xa,Susskind:2003kw}.
For inflationary phenomenology, arguably the most obvious consequences of this scenario are the possibility 
of multifield dynamics and the idea that our observable universe might have originated from a tunnelling event. Both 
of these behaviours present the alluring prospect of observable effects in the CMB.

The tunnelling process gives rise to an open universe inside the bubble of the new vacuum, whose initial evolution is 
dominated by spatial curvature \cite{Coleman:1980aw}. This leaves an imprint on the power spectrum at scales crossing 
the horizon during this period \cite{Yamamoto:1995sw,Bucher:1995ga,astro-ph/9811257,astro-ph/9901135,1105.2674} \footnote{Another possible observational consequence of this scenario is the detection 
of collisions between our bubble and other topological defects \cite{1501.05397} or bubbles 
\cite{Aguirre:2007an,Chang:2007eq,Aguirre:2009ug,Kleban:2011pg} created in the parent vacuum. This is an 
interesting possibility that is currently being actively investigated.}. If the inflationary epoch inside the bubble is 
short enough, one could hope to see this effect in the low-$\ell$ (largest observables scales) regime of the 
CMB spectrum. This possibility is heavily constrained 
by the latest Planck data \cite{1502.01589}, which sets a bound on the scale of curvature as being at least an order 
of magnitude bigger than the scale of today's horizon, implying that the scales we observe left the horizon after 
curvature domination.	

Multifield effects are potentially very important for inflation embedded in fundamental physics.
Obtaining an extended period of inflation is notoriously difficult in string theory but it is 
often the case that whatever mechanism enables one field to be sufficiently light, will tend 
to make other fields sufficiently light to be cosmologically relevant as well. This point has been emphasised 
in a large body of work but the message in Ref.~\cite{1202.4580} is particularly succinct; multifield dynamics should 
be considered the norm in string theory. 

Inflation with more than one light field can in principle give rise to a wide range of signatures, 
generally (but not exclusively) as a consequence of the superhorizon evolution of the primordial curvature perturbation, 
which in turn is a consequence of the presence of isocurvature degrees of freedom. Currently there is no observational evidence for any of these signatures, implying no evidence of multifield inflation. However it should be noted that this does not provide a confirmation against these scenarios either, since many models that exhibit multifield dynamics do not break the single-field predictions enough to be under threat from current constraints (see for instance \cite{1101.1619,1111.6646,1103.2775,1203.3792,1303.3611,1207.0317,1312.4035,1404.7496,1409.2498}).

In this work we will focus on the particularly interesting observation of a possible suppression of the power spectrum on large scales, as suggested 
by the WMAP~\cite{1212.5226} and Planck~\cite{1502.02114} data. 
There has been considerable interest in studying mechanisms for this lack of power 
\cite{astro-ph/0303636,astro-ph/0305020,
1202.6630,1309.3413,1309.4060,1403.5231,1403.7786,1404.0373,
1404.0360,1404.2278,1405.6769,1405.2012,1407.1048,1407.5816,0809.3915,0904.2518,1403.4596}, in particular as a consequence of the 
inflationary era, but its possible microphysical origin remains undisclosed. 
It was pointed out in Ref.~\cite{hep-th/0505232} that tunnelling from a metastable vacuum through a potential barrier 
might naturally generate a period of steep inflaton potential before the slow-roll plateau.  
Should we be fortunate enough that the total period of inflation is sufficiently small and,  at the same time, long enough
to avoid the constraints of curvature, we could hope to see the onset of this epoch imprinted in the CMB as some form 
of signature on large scales. Of key relevance to this paper, Bousso, Harlow and Senatore~\cite{1404.2278,1309.4060} 
emphasised that in the case of a theory consisting of a single scalar field, power suppression is a likely consequence
of this tunnelling in the landscape.

The possibility that power suppression may be a consequence 
of the string landscape is extremely exciting, as it would represent a unique door to the physics of the early universe. However there is a considerable way to go before such claims can be made. Building on the work of Ref.~\cite{1404.2278,1309.4060}, an obvious next 
question to ask in furthering this pursuit is \emph{does power suppression also occur if more than one 
field is involved in this event?}

The tunnelling process in a multifield model can be quite complicated and one might expect the initial condition 
after the transition to be quite far from the inflationary plateau region that leads to most of the observable scales 
in the CMB, the $\ell>50$ range \footnote{See \cite{BlancoPillado:2012cb} for a proposal that links the position of the fields
after the tunnelling event to the inflationary region in a multifield landscape.}.
One can expect scenarios where many moduli fields are affected by the tunnelling transition, 
potentially giving rise to a rich mass spectrum of the scalar fields involved in inflation. As an example 
of hierarchies in the mass spectrum, one can conceive that tunnelling would affect different moduli 
sectors in distinct ways. This could result, for example, in a situation where at first inflation is controlled 
by the evolution of complex structure moduli and then, as these fields settle into their minima, 
becomes dominated by a different moduli sector, like a K\"{a}hler moduli field.
Many current models of compactification 
naturally lead to this kind of hierarchies in the mass spectrum and separation in scalar sectors, indicating 
that multifield dynamics are relevant after tunnelling events (see, for example Ref.~\cite{hep-th/0610102}).
In this paper we explore how the single-field results of power suppression on 
large scales generalise in the presence of many scalar fields with such rich mass hierarchies. 

We start our discussion by reviewing a simple method to compute the power spectrum 
and the spectral index in a multifield inflationary context. 
We then review the mechanism of power suppression
due to a steepening of a single-field potential as described in Ref.~\cite{1404.2278,1309.4060}, and look at the consequences of including more than one light field in this scenario.
Our main result is that achieving power suppression 
is considerably more delicate in a multifield model than in the single-field case. In fact, a steep potential
and a fast evolution during the first few e-folds of inflation do not necessarily imply suppression of the power spectrum on large scales.
We identify strong constraints 
imposed  on initial conditions as well as inflaton potential in the region resulting in suppression. These constraints apply 
not only to the form of the potential along the inflationary trajectory
 but to all other directions in field space. This is a rare opportunity where 2-point statistics can be used to learn about multifield effects.  

If all relevant features originate purely from an inflationary era,  suppression on large scales would then  constrain multifield 
dynamics in a region where one also requires non-trivial evolution of the slow-roll parameter $\epsilon$. We 
argue that the ability to learn about the mass spectrum at such a crucial stage of inflation has significant 
implications for model building in the context of the string landscape. 

To conclude, we present a novel approach to obtain relative power suppression on large scales based entirely on superhorizon evolution of the perturbations. Here the idea is
not to suppress the power at large scales but to enhance it at smaller scales by 
continuously transferring power from isocurvature to adiabatic modes.  We present
a simple example where this can happen which could be relevant for some string theory
scenarios.

\section{Multifield inflation --- a geometrical picture} 
\label{background}

Our goal is to gain intuition about the phenomenology underlying power suppression and so to aid us in this task we will compare two methods of computing the perturbations. One approach is the transport method --- a fully numerical approach which solves for the field--field, field--momenta and momenta--momenta correlation functions, which we do not review here. We use the publicly available Mathematica code available from \href{https://transportmethod.wordpress.com/}{transportmethod.com} to perform our analyses and refer the reader to Ref.~\cite{1502.03125} for a detailed description of the method. Our second approach is a slow-roll superhorizon analysis of the perturbations which we do now review. The key benefit of this second method is that it enables one to derive a number of analytic or semi-analytic expressions which will be our primary tool in understanding the phenomenology underlying our numerical results. 

Restricting our studies to superhorizon scales and assuming the slow-roll approximations radically simplifies the computation of observable quantities. On superhorizon scales the field perturbations evolve classically in the sense that decaying solutions to the mode equations have died away. The upshot of utilising the slow-roll equations is that the evolution of all quantities of interest can be understood purely in terms of gradient flow. Hence a problem which in principle requires understanding potentially complicated operator equations, under these restrictions, reduces to a question of computing relatively simple geometrical quantities in field space. This simplification has been used in a very large body of work and is intimately related to the celebrated separate universe assumption \cite{astro-ph/0306620, astro-ph/0411220,astro-ph/9507001,astro-ph/0003278,astro-ph/0504045}. We refer the reader to Ref.~\cite{1203.2635} and references therein for a more detailed discussion. The notion of using geometrical quantities to compute observables was most heavily emphasised in \cite{1203.2635, 1005.4056, 1011.6675, 1111.0927} and it is these works which we summarise here.

Throughout this paper we work in units of $c=\hbar=1$,  
so that the reduced Planck mass reads $M_{p}^{-2}= 8 \pi G=1$, and assume the background space-time metric to be
the Friedman-Robertson-Walker with ``mostly plus" signature $(-,+,+,+)$
\be
ds^2 = -dt^2 + a(t)^2 d{\bf x}^2.
\ee

\subsection{Background equations of motion}
We will consider models describing  the inflationary dynamics of a set of $\scN$  real scalar fields, 
$\phi^i$, $i=1, \ldots, \scN$.  Restricting ourselves to theories involving at most  two space-time 
derivatives, the action can be written as
\begin{equation}
\label{action}
S = - \int d^4x\, \sqrt{-g} \Big(\ft{1}{2} G_{ij} \, \partial_\mu \phi^i \, \partial^{\mu} \phi^j + V(\phi)\Big),
\end{equation}
where  $G_{ij}= G_{ij}(\phi)$ is an arbitrary symmetric matrix and $V(\phi)$ is the scalar 
potential. The matrix $G_{ij}$ is usually interpreted  as a metric on a $\scN-$dimensional scalar 
manifold $\cM$ parametrized by the fields $\phi^i$. 
The equations of motion for the homogeneous background fields $\phi^i = \phi^i(t)$ are
\be
\label{eq:BKGEOM}
\frac{D \dot \phi^i}{dt}+3 H \dot \phi^i + G^{ij} \nabla_j V = 0, 
\ee 
where dots denote derivatives with respect to cosmological time,
$H= {{\dot a}\over a}$ is the Hubble parameter,  and we have defined 
the covariant derivative   involving the Christoffel symbols  
$\Gamma^i_{jk}$ associated to $G_{ij}$  as $D X^{i} \equiv d X^{i} + \Gamma_{jk}^i X ^j d \phi^k$.

In the following we
will focus on the evolution of the set of fields that are light enough to be part of the slow-roll dynamics. This does not preclude the existence of a heavy sector but we will assume that these other
fields do not play a significant role for the background or the perturbations. However, situations where this assumption fails can be captured using the transport method and we do include examples of this in later sections.

The slow-roll conditions in the multifield case can be defined in terms of the parameters~\footnote{We raise and lower indices  using the  metric $G_{ij}$ and its inverse $G^{ij}$.} 
\be
\label{SRcond}
\epsilon\equiv - \dot H/H^2  \ll 1; \qquad \qquad  |\mathbf{M}| \equiv |M^i_{j}| \equiv |\nabla^i \nabla_j \log V| \ll1.
\ee
Under these conditions, the equations 
of motion reduce to gradient flow equations
\begin{equation}
\label{Phi EoM slow-roll}
\phi^{i \, \prime }  =   - G^{ij} \nabla_j \log V, \
\end{equation}
where we have used the short hand notation $'\equiv \frac{d }{d N}$ to refer to derivatives 
with respect to the number of e-folds $dN = H dt$.

\subsection{Evolution of scalar perturbations}

To study fluctuations about the homogenous background we choose to work in the flat gauge such that the independent degrees of freedom are the fluctuations in the fields $\delta\phi^{i}$. These perturbations transform covariantly under a change of coordinate basis. In other words, $\delta\phi^{i}$ should be interpreted as a tangent vector rather than a coordinate displacement \cite{1101.4809}. Our task now is to derive an equation of motion for these perturbations. There are at least two ways to do this. One method is to perturb the action \eqref{action} as was done in Refs.~\cite{1208.6011,astro-ph/9507001,1502.03125}, however, given our simplifying assumptions of only considering superhorizon scales in the slow-roll regime,  there is a more direct method which is to instead perturb Eq.~\eqref{Phi EoM slow-roll}. The separate universe assumption provides an intuitive picture of why this method works.
 It states that when spatial patches are smoothed on a scale much larger than the horizon scale, the average evolution of each patch evolves according to the background equations of motion \cite{astro-ph/0306620, astro-ph/0411220,astro-ph/9507001,astro-ph/0003278,astro-ph/0504045}. The resulting picture in field space is a bundle of non-interacting trajectories, each evolving according to Eq.~\eqref{eq:BKGEOM} but subject to perturbed initial conditions. When slow-roll holds such that Eq.~\eqref{Phi EoM slow-roll} applies, this description becomes precisely analogous to geometrical optics \cite{1203.2635}. Hence, the field fluctuations $\delta\phi^{i}$ may be interpreted as Jacobi fields satisfying~\cite{1208.6011,1005.4056, 1111.0927,1203.2635}
\begin{equation}
\frac{D\boldsymbol{\delta \phi}}{dN} = - \mathbf{\tilde{M}} \cdot \boldsymbol{\delta \phi},
\label{eqPert}
\end{equation}
where
 $\mathbf{\tilde M}$ can be seen as an effective mass matrix, which encodes the couplings 
between the fields as well as a correction due to the non-trivial geometry of the scalar manifold. Denoting 
$R^i_{jkl}$ to be the components of the  Riemann tensor on $\cM$, the  matrix
$\mathbf{\tilde M}$ is 
defined as~\footnote{In the slow-roll limit we must also require that $|R_{j}^i| \ll 1$} 
\be
\tilde{\mathbf{M}}= \mathbf{M} - \frac{1}{3} \mathbf{R}\qquad \text{with} \qquad R_{j}^i \equiv R^i_{klj} \phi^{k \, \prime} \phi^{l\, \prime}.
\ee
Ultimately the principal observables of interest are the correlation functions of the primordial curvature perturbation $\zeta$. We therefore need an expression that relates field perturbations in the flat gauge to the primordial curvature perturbation in the constant density gauge \footnote{We choose to work with $\zeta$, the curvature perturbation in the constant density gauge, but we could have equally worked in the comoving gauge where the curvature perturbation is usually denoted $\cR$. Both of these quantities can be computed using cosmological perturbation theory and are known to be equal at second order on superhorizon scales up to $\mathcal{O}(k/aH)^{2}$ corrections \cite{0809.4944,astro-ph/0411463}.}. This gauge transformation can be computed using cosmological perturbation theory, however once again the separate universe assumption enables us to take a shortcut. Lyth and Rodr\'{i}guez \cite{astro-ph/0504045} showed that the separate universe assumption can be used as a practical means of computing $\zeta$ since on superhorizon scales $\zeta=\delta N$, the variation in the number of e-folds between an initial flat slice and a subsequent constant density slice. At lowest order we have
\begin{align}\label{eq:gaugetran}
\zeta(N_{\rm f}) &= \delta N \nonumber\\
&= \left(\frac{\partial N}{\partial\phi^{i}}\delta\phi^{i}\right)_{N=N_{\rm f}} \nonumber\\
&=\left(\frac{\phi'_{i}}{v^{2}}\delta\phi^{i} \right)_{N=N_{\rm f}},
\end{align}
where $v\equiv \sqrt{G_{ij}\phi^{i \, \prime }\phi^{j \, \prime }}=\sqrt{2\epsilon}$ and for clarity in discussions to follow, we have explicitly labeled the fact that all quantities are to be evaluated at the final time of interest $N_{\rm f}$.
We refer the reader to Ref.~\cite{1410.3491} for a recent discussion of this topic and detailed derivations of this expression as well as higher order expressions using both cosmological perturbation theory and the separate universe assumption.

So far we have implicitly been working on the {\it coordinate basis} for our perturbations, meaning the basis 
vectors are aligned with the original fields.
One can also look at the vector of perturbations $ \boldsymbol{ \delta \phi}$ projected onto the 
so-called \emph{kinematic basis} \cite{hep-ph/0011325,hep-ph/0107272,1005.4056, 1111.0927} defined by a set of  $\scN$ orthonormal vectors $ \mathbf{e}_a =\{ \mathbf{e}_\parallel , \mathbf{e}_\perp^\alpha \}$ satisfying 
\be
\mathbf{e_\parallel}\equiv \frac{ \boldsymbol{\phi'}}{v}, \qquad \frac{D \mathbf{e_\parallel}}{dN}\equiv Z_{21} \mathbf{e}^{(2)}_\perp, \qquad
 \mathbf{e}_a^\dag\cdot \mathbf{e}_b = \delta_{ab},
\label{basisDef}
\ee
where  $\alpha =2, ..., \scN$, and we have denoted the dot product between vectors by $\boldsymbol{a}^\dag \cdot \boldsymbol{b} \equiv G^{ij}a_i b_j$. Note that the quantity $Z_{21}$, which by definition we choose to be  non-negative, only vanishes when the background follows a geodesic trajectory, i.e. when $ \mathbf{e_\parallel}^\dag  \cdot \boldsymbol{ \nabla}\mathbf{e_\parallel} = 0$. 
The advantage of this basis is that by comparison with Eq.~\eqref{eq:gaugetran}, one can immediately identify the projection of the field perturbations along the inflationary trajectory with the primordial curvature perturbation $\zeta$. 
One can then write the following decomposition of the perturbations in the kinematic basis corresponding to the time $N=N_f$
\begin{equation}\label{eq:kinematicbasis}
  \boldsymbol{ \delta \phi} =v \, \zeta  \,\mathbf{e_{\parallel}} +  \sum_{\alpha=2}^{\scN}{\delta \phi_{\perp}^{\alpha} \mathbf{e_{\perp}^{\alpha}}}~.
\end{equation}
In this basis, the equations of motion for the superhorizon evolution of the perturbations become, \footnote{Note that after switching to the the orthonormal basis \eqref{basisDef} there is no distinction between upper and lower indices.   See Appendix A for
a more detailed explanation of these expressions.}
\bea
\label{mode-eqs}
\frac{d \zeta}{dN} &=& - 2 Z_{21} \frac{\delta \phi_\perp^{(2)}}{v},\nonumber \\
\frac{d\delta \phi_\perp^\alpha}{dN} &=& -[\boldsymbol{\tilde M}- \boldsymbol{Z}]_{\alpha \beta}\;  \delta \phi^\beta_\perp
\eea
where $(2)$ refers to the element $\alpha=2$, not to be confused with an exponent 2. We have introduced the matrix 
\be
\boldsymbol{Z}_{ab} = \mathbf{e}_a^\dag \cdot \frac{D \mathbf{e}_b}{dN},
\ee
that describes how quickly the kinematic basis vectors change along the inflationary trajectory; in other words, it expresses the turn rates of the basis vectors.

Looking at the equations of motion for the perturbations in this decomposition one can easily see two very
important points. The first equation tells us that the curvature perturbation is not constant, as it can get sourced by isocurvature modes via the first mode $\delta \phi_\perp^{(2)}$.
This will happen whenever there are turns in the trajectory in field space, or in other words, deviations from a geodesic motion, as this makes   $Z_{21}$ non zero.
The rest of the equations provide information about the evolution of 
isocurvature modes, which are controlled by the masses of the fields ${\bf \tilde M}$ as well 
as  possible turns of the inflationary trajectory.

The solution to the set of equations (\ref{eqPert}) or \eqref{mode-eqs} can be formally expressed without any loss of generality in terms of a transfer matrix $ \mathbf{T}(N,N_*)$ which describes the evolution from time  $N_*$ at horizon exit to a subsequent time $N$ 
\be
{\boldsymbol{\delta \phi}(N)\over v} =  \mathbf{T}(N,N_*)\cdot { \boldsymbol{\delta \phi_*}\over v_*}
\ee
where the subscript ($*$) represents evaluation at horizon exit. Thus, all the information about the superhorizon evolution of the perturbations 
is encoded in the transfer matrix $\mathbf{T}$. Using the decomposition Eq.~\eqref{eq:kinematicbasis}, the transfer matrix takes a particularly simple form. Setting  $N=N_f$ 
\begin{align}
\label{transfer matrix multi}
\left(\begin{array}{c} \mathcal{\zeta} \\ \boldsymbol{\frac{\delta \phi^{\alpha}_{\perp}}{v}} \end{array} \right) = & 
\left(\begin{array}{cc} 1 & \mathbf{T_{\mathcal{\zeta}\perp}} \\ 0 & \mathbf{T_{\perp \perp}} \end{array} \right)
\left(\begin{array}{c} \mathcal{\zeta}_* \\ \boldsymbol{\frac{\delta \phi^{\alpha}_{\perp}}{v_*}} \end{array} \right),
\end{align}
where the block $ \mathbf{T_{\mathcal{\zeta}\perp}}$ has the same dimensions as a  $(\scN-1)$ vector  and $\mathbf{T}_{\perp \perp}$ is an $(\scN-1 )\times(\scN-1)$ matrix  which represents the evolution of the entropy mode vector from horizon exit to the end of inflation. The matrix entry  $ \mathbf{T_{\mathcal{\zeta} \mathcal{\zeta} }}= 1$ represents the requirement that  curvature perturbations are conserved on superhorizon scales in the absence of entropy modes and  $ \mathbf{T_{\perp\mathcal{\zeta}}}=\boldsymbol{0}$
that curvature perturbations do not source entropy modes after horizon crossing~\cite{Wands:2000dp}.

In using the ``$\delta N$" approach to computing $\zeta$, so far we have been taking the flat surface and constant density surface to be infinitesimally separated. A useful alternative is to take the flat surface to be at horizon crossing
\begin{align}\label{deltaN}
\zeta(N_{\rm f}) &= \frac{\partial N}{\partial \phi_{\rm f}^{j}}\frac{\partial\phi^{j}_{\rm f}}{\partial\phi^{i}_{*}}\delta\phi^{i}_{*}\nonumber\\
&= \boldsymbol{ \nabla}^\dag N \cdot \boldsymbol{ \delta \phi}_* .
\end{align}
Written this way, all details of the superhorizon evolution are contained in the vector $\boldsymbol{ \nabla}N$. 
By direct comparison  with Eq.~\eqref{transfer matrix multi} we can find and expression for $\boldsymbol{ \nabla}N$ in terms of the components of the transfer matrix \footnote{Despite the fact that the vector $\boldsymbol{ \nabla}N$ contains information about the full superhorizon evolution, it belongs to the tangent space of $\cM$ at the point $\phi^i(N_*)$, and therefore it is decomposed in the kinematic basis corresponding to the time of horizon crossing $N=N_*$. This is to be  contrasted with Eq. \eqref{eq:kinematicbasis}, where the kinematic basis is that of $N=N_f$. This implies a slight  abuse of notation since we express the kinematic basis associated to these two different times with the same symbols $\mathbf{e}_a =\{ \mathbf{e}_\parallel , \mathbf{e}_\perp^\alpha \}$.}
\be
\boldsymbol{ \nabla} N = {1 \over v^*} (\mathbf{e}_\parallel +\mathbf{T}_{\mathcal{\zeta}\perp}).
\label{nablaN}
\ee
 We call the angle between $\boldsymbol{ \nabla}N$ and the direction of gradient flow at horizon crossing $\mathbf{e}_\parallel$  the {\it correlation angle} $\Delta_N$.
Since $\mathbf{e}_\parallel$ and $\mathbf{T}_{\mathcal{\zeta}\perp}$ are orthogonal to each other, using Eq.~\eqref{nablaN} we conclude that
\be
-\mathbf{e}_\parallel^\dag \cdot \mathbf{e}_N \equiv \cos \Delta_N = (1+T_{\mathcal{\zeta}\perp}^2)^{-1/2},
\ee
where $T_{\mathcal{\zeta}\perp} \equiv |\mathbf{T}_{\mathcal{\zeta}\perp}|$,   $\mathbf{e}_N$ is the unit vector in the direction of $\boldsymbol{ \nabla}N$, and the correlation angle is defined such that  $\Delta_N\in [0,\ft{\pi}{2}]$. This quantity, as will become clear shortly, is a very convenient  measure of the superhorizon evolution of $\zeta$.

\subsection{Two-point statistics}

We define the power spectrum of the curvature perturbation to be
\begin{equation}
\langle  \mathcal{\zeta}({\bf k}) \,   \mathcal{\zeta}({\bf k'})\rangle \equiv (2\pi)^3 \delta^3({\bf k}+{\bf k'})\, \frac{2\pi^2}{k^3}P_{\cal \zeta} .
\label{eq:powerr}
\end{equation}
To make use of expression \eqref{deltaN}, we need to specify the conditions for the perturbations at horizon crossing.
Provided the inflationary trajectory is not turning too much and slow-roll approximations hold so that $H$ is approximately constant, it is reasonable to assume all perturbations to be decoupled. Under these assumptions, the  2-point function for the field perturbations in a local frame where the fields have canonical kinetic terms, can be expressed as
\begin{equation}
\langle \delta \phi_*^i({\bf k}) \,  \delta \phi_*^j({\bf k'})\rangle = (2\pi)^3 \delta^{ij} \delta^3({\bf k}+{\bf k'})\,\frac{2\pi^2}{k^3}\left(\frac{H_{*}}{2\pi}\right)^2,
\label{eq:PBD}
\end{equation}
where a star ($*$) indicates evaluation at the comoving scale $k = aH$.


Applying the transfer matrix to the spectra at horizon crossing one immediately finds that the 
spectrum of curvature perturbations at the end of inflation   has the form~\cite{astro-ph/0205253,hep-ph/0011325,1005.4056,hep-ph/0107272}
\begin{equation}
P_{\zeta} = \left({{H_{*}}\over {2 \pi}}\right)^2 |\boldsymbol{ \nabla}N|^2= \left({{H_{*}}\over {2 \pi}}\right)^2 {1\over {2 \epsilon_{*}}} \left(1 +T_{\mathcal{\zeta}\perp}^2\right). \
\label{geometric-power}
\end{equation}
Since at horizon crossing 
$T_{\mathcal{\zeta}\perp}=0$, according to Eq.~\eqref{eq:gaugetran}, we see this expression can be simply written in terms of the correlation angle as
\be
P_{\zeta} =\frac{P^*_{\zeta}}{\cos^2 \Delta_N}.
\ee
Provided 
perturbations are small, this result relies solely on the assumption that the spectrum of perturbations is well 
described by Eq.~\eqref{eq:PBD} at horizon crossing.
Note that an important consequence of assuming the field perturbations to be uncorrelated at horizon crossing is that to leading order in the the slow-roll parameters, superhorizon evolution of the scalar perturbations always gives a positive semidefinite contribution to the power spectrum. 

As we mentioned earlier, we can see from Eq.~(\ref{mode-eqs}) that for $\cos \Delta_N \neq 1$ we need  sizeable entropy perturbations, i.e. isocurvature should not decay too fast, and we should also have mode mixing, that is, the effective mass matrix should have non-zero off diagonal terms. In the slow-roll regime the presence of mode mixing occurs whenever the inflationary trajectory deviates from the geodesic motion, or in other words, when it describes a turn in field space. In Eq.~\eqref{mode-eqs} these effects are encoded in the matrices $\boldsymbol{\tilde M}$ and $\boldsymbol{Z}$.
As we shall see in later sections these remarks  have important consequences when trying to implement large scale suppression of the power spectrum in multifield inflationary models.

We conclude this section by presenting an expression for the spectral index in this geometrical framework.
As shown in Appendix B, in the slow-roll regime the spectral index can be written in a very compact way as
\begin{equation}
\label{ns}
\frac{d \log P_\zeta}{d \log k}\equiv n_{s} -1  = -2\epsilon_* + 2  \mathbf{e}_N^\dag\cdot  \mathbf{\tilde{M}}_* \cdot \mathbf{e}_N. 
\end{equation}
This expression is the same as that found in  Ref.\cite{astro-ph/9507001}, which is the  generalisation of the one presented in Ref.~\cite{1005.4056} for two-field inflation in the slow-roll 
slow-turn regime. Note that this result implies that the spectral  tilt is determined by  two local quantities, 
$\epsilon$ and $\mathbf{\tilde{M}}$, and one non-local quantity, the unit 
vector $ \mathbf{e}_N$ which depends on the details of  the whole inflationary trajectory  between 
the time of horizon-crossing   $N_*$ and the end of inflation $N_f$, since it depends on the transfer function  $\mathbf{T}_{\mathcal{\zeta}\perp}$.
When there is no  superhorizon evolution 
the transfer matrix \eqref{transfer matrix multi} reduces to the identity matrix, implying that $\mathbf{T}_{\mathcal{\zeta}\perp}=0$
and $\mathbf{e}_N$ is parallel to the direction of the inflationary trajectory $\mathbf{e}_\parallel$. However, in general $\mathbf{e}_N$ can point in any arbitrary 
direction with $\Delta_N$ determining the relative angle between  $\mathbf{e}_N$ and 
$\mathbf{e}_\parallel$. 

\section{Power suppression in single-field inflation}

\label{sec:singleField}

To understand the idea behind power suppression in the presence of a steepening of the potential, let us briefly discuss the 
 case of single-field models. In this case, assuming slow-roll is a good approximation at horizon crossing, the power spectrum for the scale $k_*=a_*H_*$ that crosses the horizon at $N_*$ is given by
 \begin{equation}
P_\zeta(k_*) = {{H_{*}^2}\over {8 \pi^2 \epsilon_{*} }} . \
\label{singlePower}
\end{equation}
Since during inflation the Hubble parameter is approximately constant, to realise $P_\zeta(k_1) <P_\zeta(k_2)$, with $k_1<k_2$, requires that $\epsilon(N_1) > \epsilon (N_2) $. In other words, power suppression between $k_1$ and $k_2$ can only occur if the slow-roll parameter $\epsilon$ is decreasing between the times of horizon exit of these scales, $N_1$ and $N_2$ respectively. This implies that the potential should be \emph{steeper} between $N_1$ and $N_2$ than in the subsequent inflationary evolution.

\begin{figure}
  \centering
  \includegraphics[width=.49\textwidth]{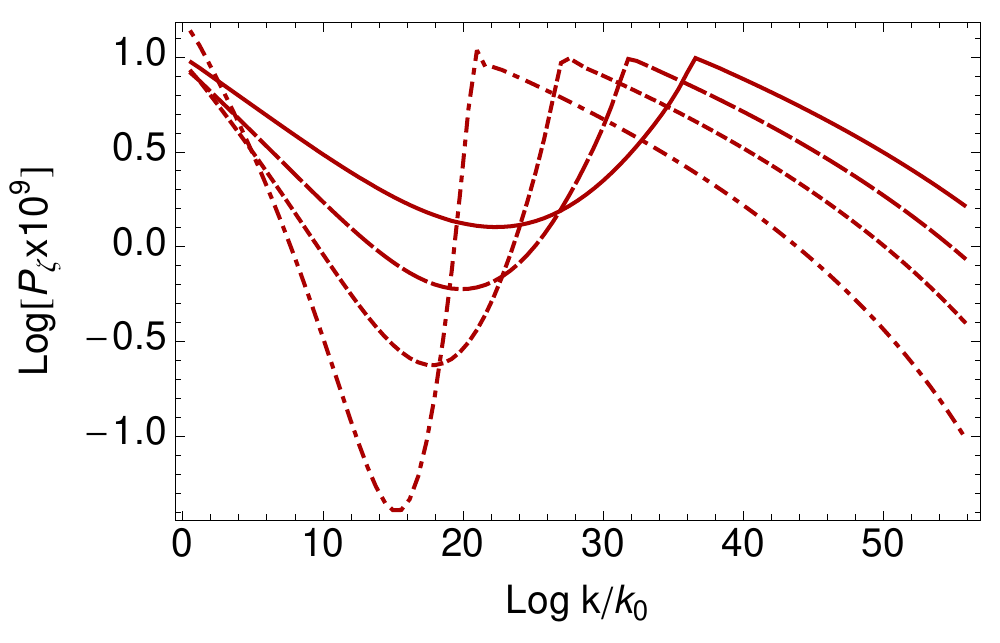}
\includegraphics[width=.49\textwidth]{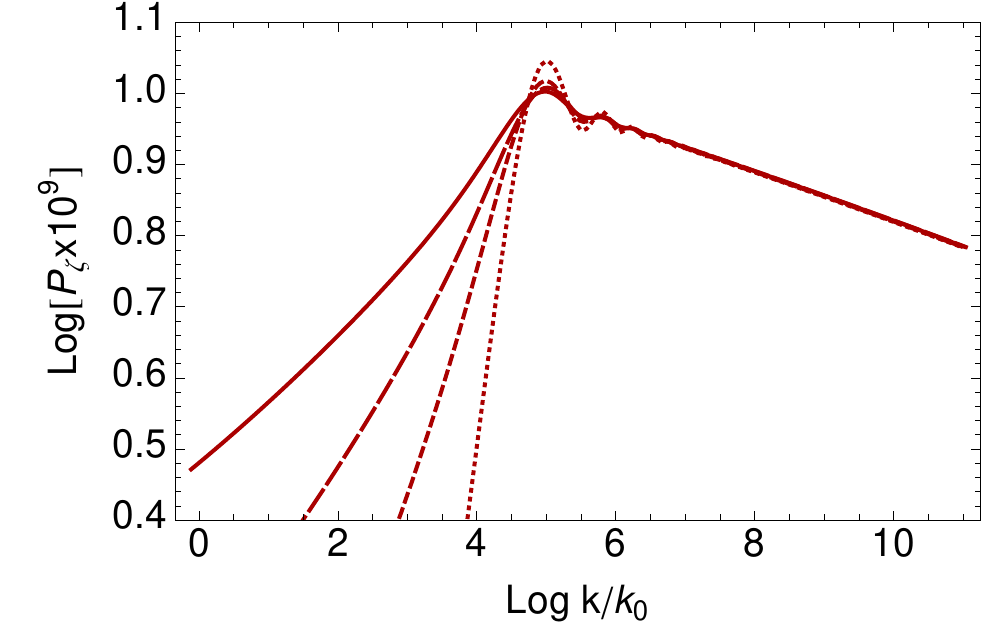}
   \caption{Power spectrum of the  model \eqref{eq:Vofphi1} with parameters $\epsilon_{\rm S}=0.005$, 
$\Lambda^4 = 3.5 \times 10^{-9}$ and different values of the mass parameter $m_1^2$. Both plots correspond to  $m_{1}^{2}=0.07, \, 0.1, \, 0.15, \, 0.3$ represented by a solid, 
long-dashed, short-dashed and  dotted lines respectively.
LEFT:  We have set  $k_0$  to be the scale $k$ that exits the horizon at the time the $N_ e=0$. 
 RIGHT: In order to make contact with observations it is more convenient to fix  $k_0$  so that the scale $k$ that exits the horizon at the time the transition occurs $\phi(N_c=4)=\phi_c$ satisfies $\log k/k_0 \approx 4$. 
   }\label{fig:Pofk_1d}
\end{figure}

The simple example  proposed in Refs.~\cite{1404.2278,1309.4060}  illustrates this  idea well. The potential has the general form
\begin{equation}\label{eq:Vofphi1}
V(\phi)= \Lambda^{4}(V_{\rm S}(\phi) + V_{\rm R}(\phi)),
\end{equation}
where $V_{\rm S}(\phi)$ is the ``slow" part of the potential, modelled as a Taylor expansion 
with coefficients chosen to match observational constraints 
\begin{equation}\label{eq:Vs}
V_{\rm S} = 1-\sqrt{2 \epsilon_{\rm S}} \phi.
\end{equation}
The mass scale $\Lambda$ is fixed by {\sc{cobe}} normalisation \cite{astro-ph/9601067} and makes $V_{\rm S}$ and $V_{\rm R}$ 
(and in particular $\phi$) dimensionless.
The early stages of inflation are dominated by a brief period of  ``rapid" evolution on a steeper potential $V_{\rm R}$, that should 
be steep enough to provide the desired power suppression. According to Refs.~\cite{1404.2278,1309.4060}, these requirements 
are satisfied by a quadratic potential like 
\begin{equation}
V_{\rm R} = \Theta(\phi_{\rm c}-\phi)\frac{1}{2}m_{1}^{2}(\phi_{\rm c}-\phi)^{2}
\label{eq:VR}
\end{equation}
where $\phi_c$ can be chosen to produce power suppression at $\ell \lesssim 50$, as  we do for example in the right plot of Fig.  \ref{fig:Pofk_1d}. 

The key observation made in Ref.~\cite{1309.4060} is that for the model given by Eq.~\eqref{eq:Vofphi1} one can obtain an approximate 
expression for this power spectrum, when $m_{1}^{2} \ll 1$, of the form
\begin{equation}
P_{\zeta}\approx \left \{1-2 m_{1}^{2} \frac{V_{\rm R,\phi}}{V_{\rm S , \phi}}\right \}^{*}P_{\zeta}^{\rm S},
\end{equation}
where $P_{\zeta}^{\rm S}$ is the power spectrum that would be obtained by considering $V_{\rm S}$ only. Importantly this expression 
shows that the effect of $V_{\rm R}$ is always to suppress the power spectrum. This is intuitive, as provided Eq.~\eqref{singlePower} 
holds, one sees that the effect of the ``rapid" phase manifests as increasing the magnitude of $\epsilon$. 

\begin{figure}[t]
  \centering
  \includegraphics[width=.49\textwidth]{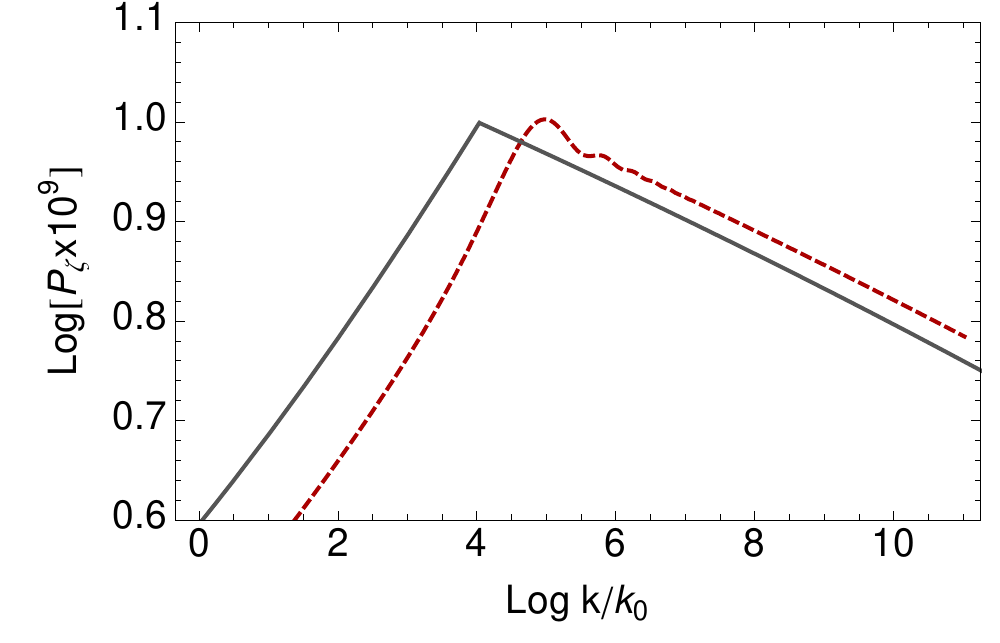}
\includegraphics[width=.49\textwidth]{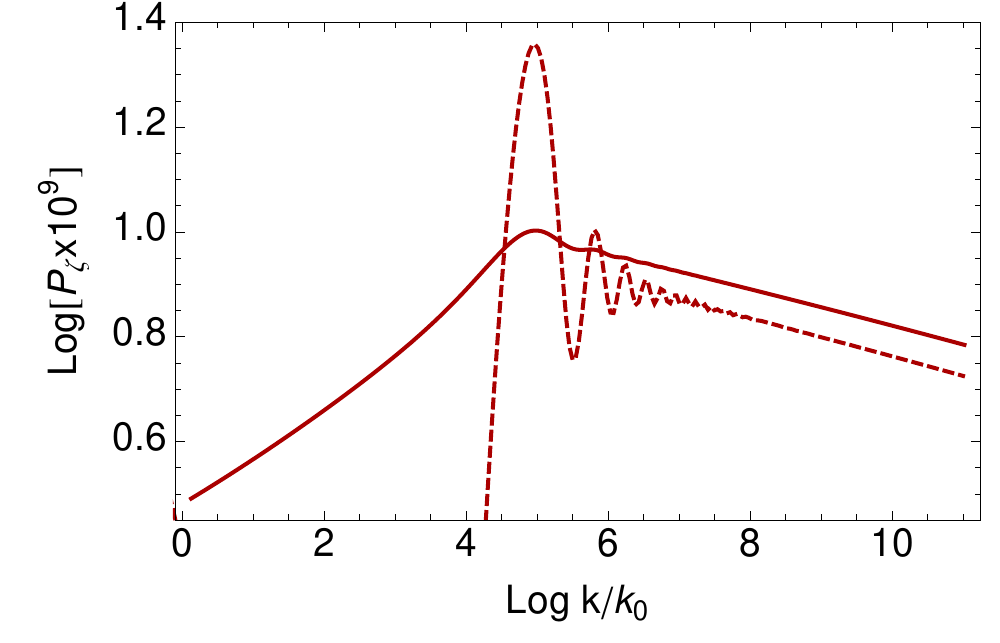}
   \caption{Power spectrum of the  model \eqref{eq:Vofphi1} with parameters $\epsilon_{\rm S}=0.005$ and 
$\Lambda^4 = 3.5 \times 10^{-9}$. We have set  $k_0$  so that the scale $k$ that exits the horizon at the time the 
transition occurs $\phi(N_c=4)=\phi_c$ satisfies $\log k/k_0 \approx 4$.  LEFT: Power spectrum for the  mass parameter $m_1^2=0.07$. The grey solid line represents the spectrum obtained from the slow-roll approximation 
\eqref{singlePower} and the red solid  line results from a full numerical simulation. RIGHT: Power spectrum 
obtained from a full numerical simulation  with mass parameters 
$m_1^2=0.07$ (solid line) and $m_1^2=2.4$ (dashed line).} \label{fig:Pofk_wiggles}
\label{fig:massDep}
\end{figure}

 To understand more accurately the power suppression induced by the presence of $V_{\rm R}$,  one should take into account the 
deviations from the slow-roll 
approximation or from \eqref{eq:PBD} induced by the period of rapid evolution. For this we performed a full 
 numerical evolution of the background and two-point correlation functions of the 
fluctuations  (which does not rely on slow-roll approximations) \cite{1502.03125} \footnote{Our numerical approach is the non-slow-roll version of the Transport method. We refer the reader to Refs.~\cite{1502.03125, 0911.3550,1008.3159,1203.2635,1302.3842} for more information on this approach.}. For simplicity, and consistency
 with Refs.~\cite{1404.2278,1309.4060}, curvature effects were completely ignored in the evolution equations. We believe that given 
the current constraints on spatial curvature this approximation should not impact on our main results, which refer to scales leaving the 
horizon after the curvature bound has been reached. In addition, if our observable universe tunnelled from a parent metastable vacuum, 
and inflation started earlier than the horizon-crossing time for the largest observable scale in order to dilute curvature artefacts, it seems 
reasonable to assume a Bunch--Davies state as our initial conditions at this time.  This is the starting point of our numerical evolution 
which we define to be at $N_e=0$ \footnote{The
initial state after tunnelling is known to be affected by the parent vacuum as well as other effects associated to the
nucleation process. Our assumption is that these effects are small for the cases studied in this paper. This point has recently 
been investigated in \cite{1407.5816}. Their conclusions indicate that this would be a small effect.}. The results of the simulations 
are displayed in 
Figs.~\ref{fig:Pofk_1d}  and~\ref{fig:Pofk_wiggles}. 

In Fig. \ref{fig:Pofk_1d} we have displayed how the power spectrum varies for different choices of the mass parameter $m_1^2$ which 
determines the steepening, while the other two parameters in the potential, $\Lambda$ and $\epsilon$, are kept fixed for simplicity. The 
spectrum in these plots interpolate between the spectrum of quadratic inflation for large scales $k \sim k_0$ and the one of linear inflation 
for small scales.  The  transition point corresponds roughly to the point where the spectral tilt $n_s-1$ changes abruptly
from  positive to negative.  The figure shows that, as the ``rapid''  region of the potential becomes steeper (for increasing values 
of $m_1^2$), the power suppression in the 4 e-folds prior to the transition becomes more pronounced, and at the same time the 
range of scales which experience power suppression decreases.

  In order to identify  which models are capable of producing power suppression on large scales, it is useful to 
  observe that  the suppressed spectra necessarily have a positive (blue) tilt i.e.  $n_s-1>0$, at some point during 
  the first $4$ e-folds of the observable inflation. We therefore use the deviations of the spectral index on large scales as a marker 
for  the existence of power suppression. As we discussed in the previous section, the general expression for the spectral index in 
the multifield case is complicated by the fact that it involves non-local terms. However, in the case of single-field inflation, things
 become much simpler since the vector $ \mathbf{e}_N$ necessarily coincides with the direction of the inflaton $\mathbf{e}_\parallel$,  
and therefore the expression for the spectral index to leading order in the slow-roll parameters, Eq.~(\ref{ns}), reduces to the usual formula 
\be
\mathbf{e}_\parallel^\dag\cdot  \mathbf{\tilde{M}}\cdot   \mathbf{e}_\parallel= - 2 \epsilon+ \eta \qquad \Longrightarrow \qquad n_{s} -1=  -6\epsilon + 2 \eta.
\label{nsCond0}
\ee
Here  $\eta$ is the  second slow-roll parameter of single-field inflation given  by  $\eta \equiv V_{,\phi\phi} / V$. 
Using this expression for the spectral index we can translate the blue tilt condition on the  spectrum  into a  constraint on the inflationary 
potential.  According to Eq.~\eqref{nsCond0}, a blue tilt for the scale $k_*$ occurs  whenever\be
 \eta > 3 \epsilon \qquad \Longrightarrow \qquad V_{,\phi\phi} V > {3\over 2} V^{ 2}_{,\phi}.
\label{nsCond1}
\ee
We expect the slow-roll approximation to be in good agreement with the numerical result for scales exiting the horizon well away 
from the transition between the rapid and slow parts of the potential. However scales leaving the horizon during the transition between 
these two periods of inflation can experience effects that cannot be captured by the slow-roll description. One effect is the mixing of 
mode functions, which gives rise to oscillations in $P_\zeta(k)$ as can be seen  in Fig.~\ref{fig:Pofk_wiggles}. This effect is well studied 
and we refer the reader to Refs.~\cite{astro-ph/0303636,1405.2012,1407.1048} for more detailed discussion
\footnote{ Recently, in 
Ref.~\cite{1407.1048}, a study of the implications of different equations of state at the beginning of inflation was performed (for related work see \cite{Burgess:2005sb, Allahverdi:2007ts}). In this 
paper~\cite{1407.1048}, authors considered an instantaneous change between pre- and inflationary eras. We believe that slow approaches to the 
inflationary attractor might also have interesting phenomenological consequences.}.

 In the left 
plot of Fig.~\ref{fig:Pofk_wiggles} we compare the numerically obtained non-slow-roll result with the slow-roll approximation which 
we obtained  solving the gradient flow equation Eq.~\eqref{Phi EoM slow-roll} and  assuming slow-roll at horizon crossing, 
Eq.~\eqref{singlePower}. When the mass $m_{1}$ is small, as for the model parameters discussed in Refs.~\cite{1404.2278,1309.4060}, then the non-slow-roll correction is small.  However, it is noteworthy that increasing the mass $m_{1}$ does not simply reduce the range of scales over which power suppression 
occurs; we also see that the oscillations  became  large.  An example of this is shown on the right plot of Fig.~\ref{fig:Pofk_wiggles}, 
where we compare the spectra obtained with $m_1^2=0.07$ and $m_1^2=2.4$.  If the potential is too steep the resulting power spectrum 
could be ruled out by observations, imposing important constraints on this model as a viable method to obtain power suppression. 

\section{Power suppression in multifield inflation}
\label{sec:suppresionMulti}

The construction of multifield inflationary models which display  suppression of the power spectrum at large scales  
is more subtle than in single-field models. Consider the ratio of the 
power spectrum $P_\zeta(k)$ at two different scales $k_1 <k_2$ which exit the horizon   at the times $N_1$ and $N_2$ 
respectively. In order for the power spectrum to be suppressed at the  scale $k_1$ with respect to $k_2$ we need to have
\be
\label{constraint}
\frac{P_\zeta(k_1)}{P_\zeta(k_2)} = \frac{P_{\zeta}^*(k_1)}{P_{\zeta}^*(k_2)}\, \left(\frac{\cos^2\Delta_{N_2}}{\cos^2\Delta_{N_1}}\right) <1.
\ee
In other words, this ratio depends on the relative amount of superhorizon evolution experienced by scales $k_1$ and $k_2$, 
which in turn depends on 
the relative size of the corresponding correlation angles $\Delta_{N_*}$.
 Recall that  
$\Delta_{N_*}\in [0,\pi/2]$ is the angle between the gradient of the number of e-folds, 
parallel to $\mathbf{e}_N$,  and the inflaton direction at the time $N_*$, 
$\mathbf{e}_{\parallel}$.  For a  given ratio $P_{\zeta}^*(k_1)/P_{\zeta}^*(k_2)$  at horizon 
exit, we can find two different situations:

\begin{itemize}
\item {\bf $\Delta_{N_2} \le \Delta_{N_1}$:} When the correlation angle is smaller for the smaller scale the superhorizon evolution acts as to  \emph{reduce}
 the power suppression existing at horizon exit
\be
\frac{P_\zeta(k_1)}{P_\zeta(k_2)} \ge  \frac{P_{\zeta}^*(k_1)}{P_{\zeta}^*(k_2)}.
\ee

As a consequence of this, to have power suppression at the end of inflation in this scenario, it is not sufficient to have it 
 present at horizon crossing.

\item {\bf $\Delta_{N_2^*} >\Delta_{N_1^*}$:} When the correlation angle is smaller for the larger scale 
 the superhorizon 
evolution acts as to  \emph{enhance} the power suppression existing at horizon exit 
\be
\frac{P_\zeta(k_1)}{P_\zeta(k_2)} <  \frac{P_{\zeta}^*(k_1)}{P_{\zeta}^*(k_2)}.
\ee
In this situation power suppression might be present at the end of inflation even when it is absent at horizon crossing, 
i.e. $P_{\zeta}^*(k_1) \approx P_{\zeta}^*(k_2)$. 

\end{itemize}

For an inflationary model to be predictive, inflation must reach the adiabatic limit before the end 
of inflation. In scenarios where this is the case, it seems reasonable to expect  that scales exiting the horizon at later times experience progressively less 
superhorizon evolution. From this perspective, the first situation seems somehow more typical than the second one. The second situation can occur, for example,  when a very massive field suddenly becomes light at some point between $N_1$ and $N_2$. In that case, low power at large scales arises because of a power enhancement at small scales. In the next section we will show a concrete example of the first situation, and in \S~\ref{SHsuppression}  we will present an example of the second.

\paragraph{Analysis of the spectral index:}

Similarly to the  case of single-field inflation, a useful way to translate the conditions for power suppression into constraints on the form of the potential is to look at the evolution of the spectral index.
As we argued above, the presence of suppression at large scales requires  a period where the tilt transitions to blue. In the multifield case a blue tilt for the scale $k_*$ occurs  whenever 
\begin{equation}
\label{blueCondition}
  \mathbf{e}_N^\dag \cdot \mathbf{\tilde{M}}_* \cdot  \mathbf{e}_N >  \epsilon_* , 
\end{equation}
in other words, whenever the projection of the effective mass matrix $\mathbf{\tilde M}_*$ along the vector $\mathbf{e}_N$ is larger 
than the slow-roll parameter $\epsilon_*$.  In general the vector $\mathbf{e}_N$ will not point in the direction of the inflationary trajectory and 
therefore, in contrast to the single-field case, the condition above is unrelated   to the steepness of the potential.
In fact, the 
steepening of the potential is neither necessary nor sufficient to guarantee a period of blue tilt in the large scale power spectrum.

To understand the constraints on the potential around 
the inflationary trajectory, it is convenient to express Eq.~\eqref{blueCondition} in terms of the Hessian of the scalar potential, 
rather than $\mathbf{\tilde M}$ which is related to the Hessian of $\log V$:
\begin{equation}
\label{blueCondition2}
  \mathbf{e}_N^\dag \cdot  \mathbf{\tilde H}_* \cdot \mathbf{e}_N > (1+2\cos^2\Delta_{N_*})\,  V_*\, \epsilon_* , \qquad \text{where} 
\qquad \tilde H_{j}^i \equiv \nabla^{i}\nabla_j  V - \frac{1}{3} R_{j}^i.
\end{equation}
When the field space metric is flat everywhere  the curvature term in the definition of $\mathbf{\tilde  H}$ is absent and it 
reduces to the Hessian of $V$.  
Since the diagonal elements of any hermitian matrix are always bounded 
above by the largest eigenvalue, this condition implies that the largest eigenvalue of $\mathbf{\tilde H}_*$ must  be larger
 than at least $V_*\epsilon_*$. 
Therefore, a necessary but not sufficient condition to have a blue tilt in the spectrum at a given scale $k_*$ is that the largest mass of 
the system evaluated at horizon exit satisfies 
\be
m^2_*|_{\text{max}} > V_*\, \epsilon_* .
\label{necessBlue}
\ee
 In the same way, since  diagonal
 elements of a hermitian matrix are bounded below by the minimum eigenvalue, we find the sufficient but not necessary condition for blue tilt at $k_*$ 
\be
m^2_*|_{\text{min}} > 3 V_*\, \epsilon_*. 
\ee
In other words, when this condition is satisfied the power spectrum will always be blue-tilted at $k^*$, irrespective of the superhorizon evolution experienced by the mode after horizon exit.     
\section{Example I: plateau preceded by quadratic inflation}

In this example we consider the situation where {\bf $\Delta_{N_2} \le \Delta_{N_1}$}. We discuss a two-field model where the fields have canonical kinetic terms $G_{ij}=\delta_{ij}$, and we define the  scalar potential to be
\begin{equation}\label{eq:Vofphi2}
V(\phi_{1},\phi_{2})= \Lambda^{4} \Big [V_{\rm S}(\phi_{1}) +V_{\rm R1}(\phi_{1}) + V_{\rm R2}(\phi_{2})\Big ].
\end{equation}
Here $V_{\rm S}$ and $V_{\rm R1}$ remain the same as in Eqs.~\eqref{eq:Vs} and \eqref{eq:VR}, and $V_{\rm R2}$ is given by
\begin{equation}
V_{\rm R2}=\frac{1}{2}m_{2}^{2}\phi_{2}^{2}.
\end{equation}
This illustrates the picture where  a mass hierarchy is present in the scalar sector resulting in non-trivial dynamics before the inflationary plateau.  This is the typical situation after a tunnelling event where the initial conditions for inflation are generically far away from the inflationary region of the potential.

The first thing we can tell about this model is that in order to have power suppression, according to the necessary condition \eqref{necessBlue}, we require that during some period 
\be
 \max\{m_1^2, m_2^2\}  >  \frac{V(\phi_c) \, \epsilon_S }{\Lambda^4},
\label{necessBlue2}
\ee
where we have also used that in the steep region of the potential $V \ge V(\phi_c)$ and $\epsilon \ge \epsilon_S$. If this 
condition is not satisfied, no slow-roll inflationary evolution will lead to a blue-tilted power spectrum, and therefore we can 
not have  power suppression on large scales.

Conveniently, this potential belongs to the sum-separable class 
\begin{equation}
V(\phi_{1},\ldots,\phi_{\scN})=\sum_{i}V_{i}(\phi_{i})
\label{separableV}
\end{equation}
for which it is possible to calculate analytic expressions for observable quantities, 
provided the background evolution is well approximated by the slow-roll equations and the kinetic terms are canonical. 
Specifically, the vector $\boldsymbol{\nabla} N$ in these models can be written as \cite{astro-ph/0603799}
\begin{equation}
\label{sepDN}
(\boldsymbol{\nabla} N)_i|_{N_f} = \frac{V_{i \, *}+Z_i|_{N_f}}{V_{,i \, *}}
\end{equation}
where $Z_i$ are a set of functions that become constant in the adiabatic limit. Without loss of generality we can always rearrange the constant terms of the summation $V=\sum_i V_i $ such that $Z_i|_{N_f}=0$ \footnote{If a field  reaches  the  minimum of its own potential during inflation $V_{,i} |_{\phi^i_0}=0$, it is always possible to redefine the potential such that $V_i|_{\phi^i_0}=0$ also vanishes at that point. It can be shown that this choice  implies that the corresponding   $Z_i$ has to be zero at  $\phi^i_0$  (see \cite{astro-ph/0603799, astro-ph/0610296}). }.
Thus,  using Eq.~\eqref{geometric-power}, we have that
\begin{equation}
\label{eq:HCA}
P_\zeta =  \left({{H_{*}}\over {2 \pi}}\right)^2\sum_{i} \left(\frac{V_{i}}{V_{,i}}\right)^2_*=P_\zeta^* \left\{ 2\epsilon_*\sum_{i} \left(\frac{V_{i}}{V_{,i}}\right)^2_*\right\}
\end{equation}
which depends exclusively on quantities evaluated at horizon crossing, even though it encodes all superhorizon evolution of the power spectrum. 
The term in curly brackets can be thought of as the result of transferring power from isocurvature perturbations to adiabatic perturbations on superhorizon scales. 
Using this expression and the condition for power suppression Eq.~\eqref{constraint}, we can immediately find constraints on the potential. The requirement for power suppression is not a period of inflation with relatively larger $\epsilon$ (like in single-field inflation), but a period where the summation $\sum_i\left(V_i/V_{,i}\right)^{2}$ is smaller\footnote{$N.B$ the explicit $\epsilon$ factor in Eq.~\eqref{eq:HCA} cancels with the term in the denominator of $P_\zeta^*$.}, which is a significantly more intricate condition. 
\begin{figure}[t]
  \centering
  \includegraphics[width=.65\textwidth]{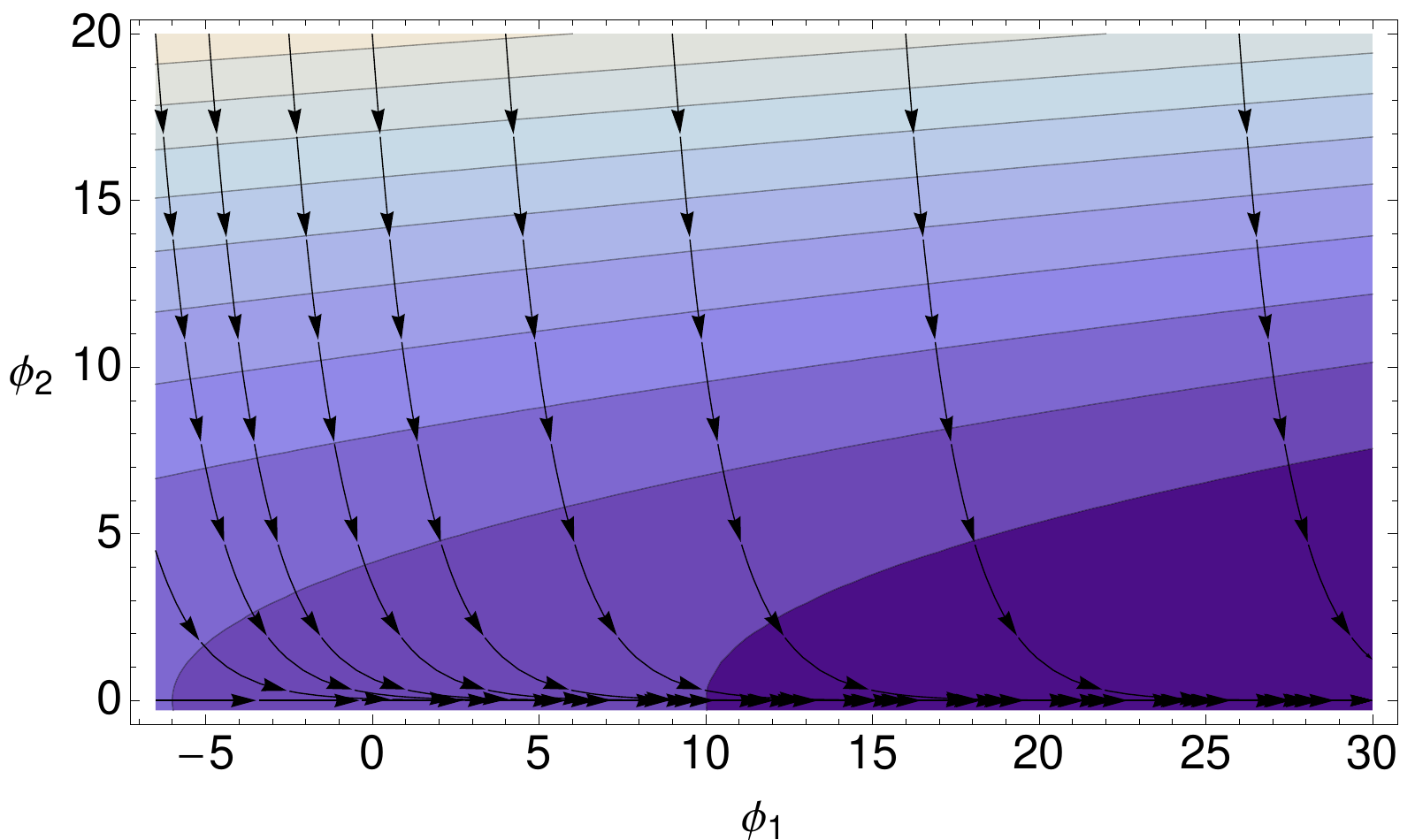}
  \caption{Scenario One $(m_{1}=0)$. We show several different inflationary trajectories 
  on the $\phi_1 - \phi_2$ plane. None of the trajectories have power suppression at 
  large scales even though it is present at horizon crossing. }
\label{scen1:trajectories}
\end{figure}

To better understand how this translates to constraints on initial conditions it is convenient to express Eq.\eqref{eq:HCA} in terms of the correlation angle. Remembering that $\Delta_N$ is the angle between $\boldsymbol{\nabla} V$ and $\boldsymbol{\nabla} N$
\be
\boldsymbol{\nabla} V = \Lambda^4(V_{S,1} +  V_{R1,1} \, ,\,  m_ 2^2 \, \phi_2),  \qquad  
\boldsymbol{\nabla} N =\left(\frac{V_{S} +  V_{R1} }{V_{S,1} +  V_{R1,1} }\, , \, \frac{\phi_2}{2} \right)
\ee
and that  $|\boldsymbol{\nabla} V | = \sqrt{2 \epsilon} V$, we obtain
\be
\cos \Delta_N =\frac{1}{\sqrt{2 \epsilon} \, |\boldsymbol{\nabla} N|}\le \frac{\sqrt{2/\epsilon}}{|\phi_2|}.
\label{SHbound}
\ee
When the system is in its adiabatic limit
the two directions coincide and $\Delta_N=0$.  Conversely, for large values of $\phi_ 2^2$ the correlation angle is close to $\pi/2$, implying that in this regime the perturbations will experience a large amount of superhorizon evolution. 
 Using the previous result it is straightforward to derive an upper bound for the
 magnitude of power suppression in the multifield case
\be
1>\frac{P_\zeta(k_1)}{P_\zeta(k_2)}
 \ge  \frac{\epsilon |\phi_2|^2}{2}\Big|_{N_1}\,  \cos^2 \Delta_{N_2}. 
\label{boundPhi2}
\ee
For an analysis of initial conditions, it is convenient to make a change of variables in field space such that
\be
\phi_1 =  \phi_c  +  \frac{\sqrt{2 \epsilon_S}}{m_1^2}  +  R \cos \theta, \qquad  \phi_2 =  R \sin \theta.
\ee  
and start the trajectories always at the same height, from a fixed value of the scalar potential $V|_{N_1}=V_0$. These initial conditions define a curve in field space which is characterised by a function $R = R(\theta)$. 
Expressing $\phi_2$ and $\epsilon$ in terms of $V_0$ and $\theta$, and keeping only the leading order terms in the slow-roll parameter $\epsilon$,  we can rewrite Eq.~\eqref{boundPhi2} as 
\be
1 \ge  \frac{\Delta V_0 \,\epsilon_{N_2}\cos^2 \Delta_{N_2}}{\Lambda^4\, (m_1^2 \cot^2\theta_{N_1} + m_ 2^2 ) },
\ee
where $\Delta V_0=V_0-\Lambda^4$ is the initial height above the inflationary plateau. As we require our model to approach the adiabatic limit in order to be predictive, for simplicity we will assume that this limit is reached shortly after the steep region of the potential, such that  $\Delta_{N_2}=0$. Moreover, we will also suppose that at this point the inflaton is rolling down the inflationary plateau where  $\epsilon_{N_2}=\epsilon_S$. In this case, the condition for power suppression can be expressed in terms of a maximum value of the initial angular direction, $\theta|_{\text{max}} =\theta_c$, above which power suppression cannot be realised:
\be
\tan^2 \theta_ c \equiv \frac{  \Lambda^4 m_1^2}{ \epsilon_S \, \Delta V_0 -   \Lambda^4 m_2^2  }.
\label{thetaC}
\ee

\paragraph{Scenario One $(m_{1}=0):$} The effects of isocurvature are especially evident in the case $m_1=0$. As argued 
in \S~\ref{sec:suppresionMulti}, in order to have power suppression at horizon crossing  the inflationary trajectory 
needs to experience a steepening,  and in this case the rapid evolution occurs in a different field direction to the slow phase, 
i.e. along $\phi_2$.  This na\"ively looks like a small change from the single-field scenario since one can still arrange for  the trajectory to follow very similar values of the potential thus resulting in a similar profile for $\epsilon$. However the role of 
multifield effects turns out to be very important -- when the superhorizon evolution is taken into account the power suppression disappears, completely. This follows directly from Eq.~\eqref{thetaC}, as the critical angle vanishes and therefore there are no initial conditions for which one can realise power suppression.

\begin{figure}[t]
  \centering
  \includegraphics[width=1.\textwidth]{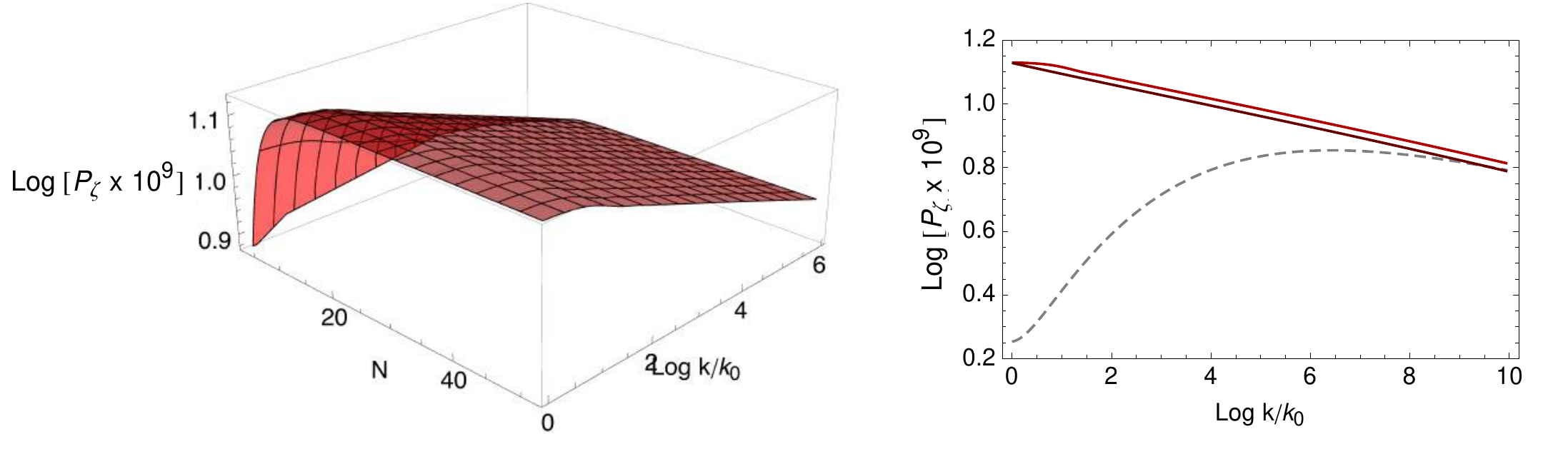}
  \caption{Superhorizon evolution of the power spectrum for scenario one --- the two-field model given by Eq.~\eqref{eq:Vofphi2} 
where $m_{1}^{2}=0$ and $m_{2}^{2}=0.2$. LEFT:  $P_\zeta(k)$ changing from horizon exit until the end of 
inflation. RIGHT:  comparison between $P_\zeta$ evaluated at horizon crossing  (dashed grey) and at the end 
of inflation; lighter red is the numerical result and dark red is the analytic expression given by Eq.~\eqref{eq:HCA}.
  We see that initially, at $N=0$, there is significant power suppression on large scales (small $\log (k/k_{0})$) but soon after 
the largest scale leave the horizon, the inflationary trajectory goes through a turn, thereby giving rise to superhorizon 
evolution which acts to erase any information of this initial state. $k_0$, $\Lambda$ and $\epsilon_{\rm S}$ are defined 
as in Fig.~\ref{fig:massDep}. }\label{fig:3dPofNk}
\end{figure} 

 The evolution of the power spectrum on each scale from horizon exit up to the end of inflation is shown in  
Fig.~\ref{fig:3dPofNk}. Around horizon crossing, one sees significant power suppression on large scales (for $\log k /k_{0} \lesssim 3$) 
but soon after these scales leave the horizon, the field-space trajectory goes through a turn (see Fig. \ref{scen1:trajectories}), 
causing the isocurvature perturbation to source the adiabatic perturbation and hence $P_{\zeta}$ evolves in such a way as to 
erase any information about the steepening of the potential that was there initially. 

The change in the spectral tilt displayed in the right plot of Fig. \ref{fig:3dPofNk}  can be understood easily from the multifield 
expression of the spectral index Eq.~\eqref{ns}.
Consider a trajectory starting at some large value of $\phi_2$ --- initially $ \mathbf{e}_{\parallel}$ points roughly 
along the $\phi_2$   direction and $ \mathbf{e}_{\perp}$ points along the $\phi_1$ direction (see Fig. (\ref{scen1:trajectories})).   
If there was no superhorizon evolution, the spectral index would then reduce to 
\be
n_s -1 \approx -6 \epsilon_* + \frac{2 \Lambda^4 m_2^2}{V}, 
\ee
implying that at a certain height of the potential $V$ we could have a blue-tilted spectrum by choosing $m_2^2$ to be 
sufficiently large.
%
This would be the na\"ive  conclusion one would get by studying the spectrum at horizon crossing. However, making superhorizon evolution into account, from the bound \eqref{SHbound}, we see that at large $\phi_2$ the correlation angle should be close to 
$\pi/2$, implying that $ \mathbf{e}_{N}$ points instead along the direction of $\phi_1$. As a consequence, using Eq.~\eqref{ns} with $m_1^2=0$, we find that the spectral index is smaller than one
\be
n_s -1 \approx -2 \epsilon_*, 
\end{equation}
\textit{i.e.} the spectrum is  red-tilted at the end of inflation. This comparison is shown explicitly in the right plot of Fig. \ref{fig:3dPofNk}. 

\begin{figure}
  \centering
  \includegraphics[width=.48\textwidth]{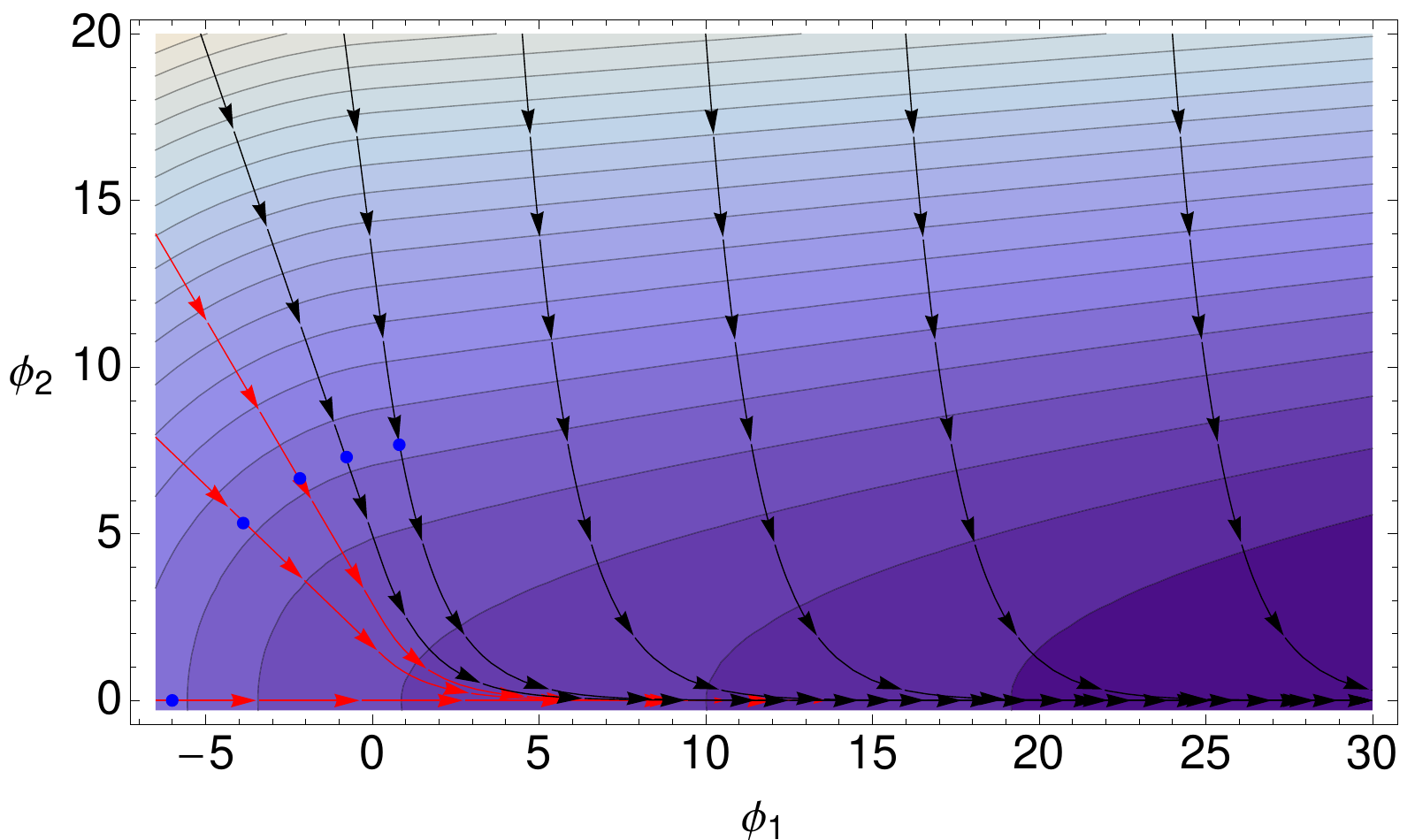}
  \hfill 
  \includegraphics[width=.45\textwidth]{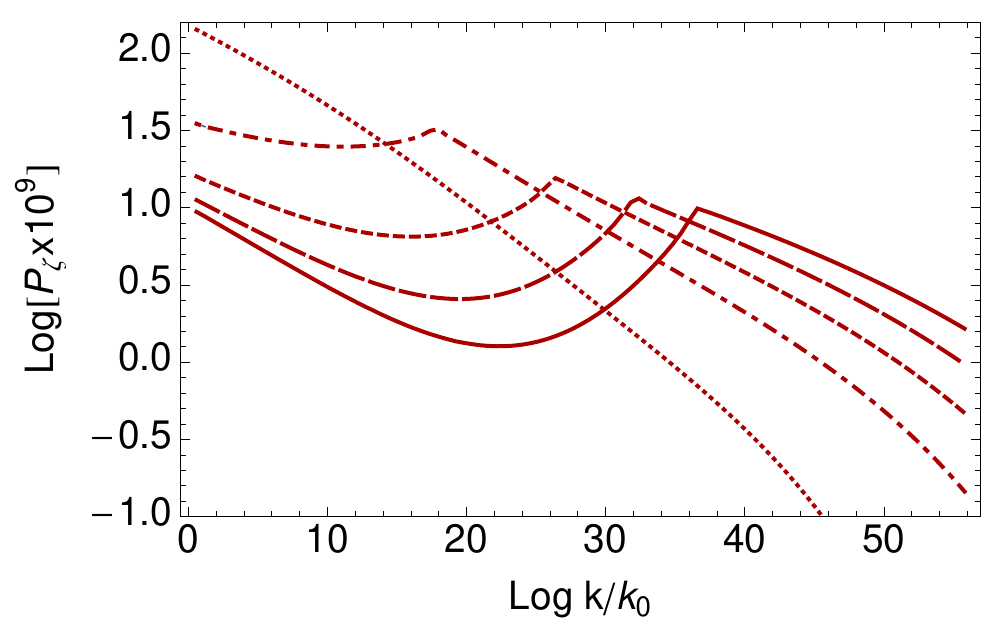} 
   \includegraphics[width=.48\textwidth]{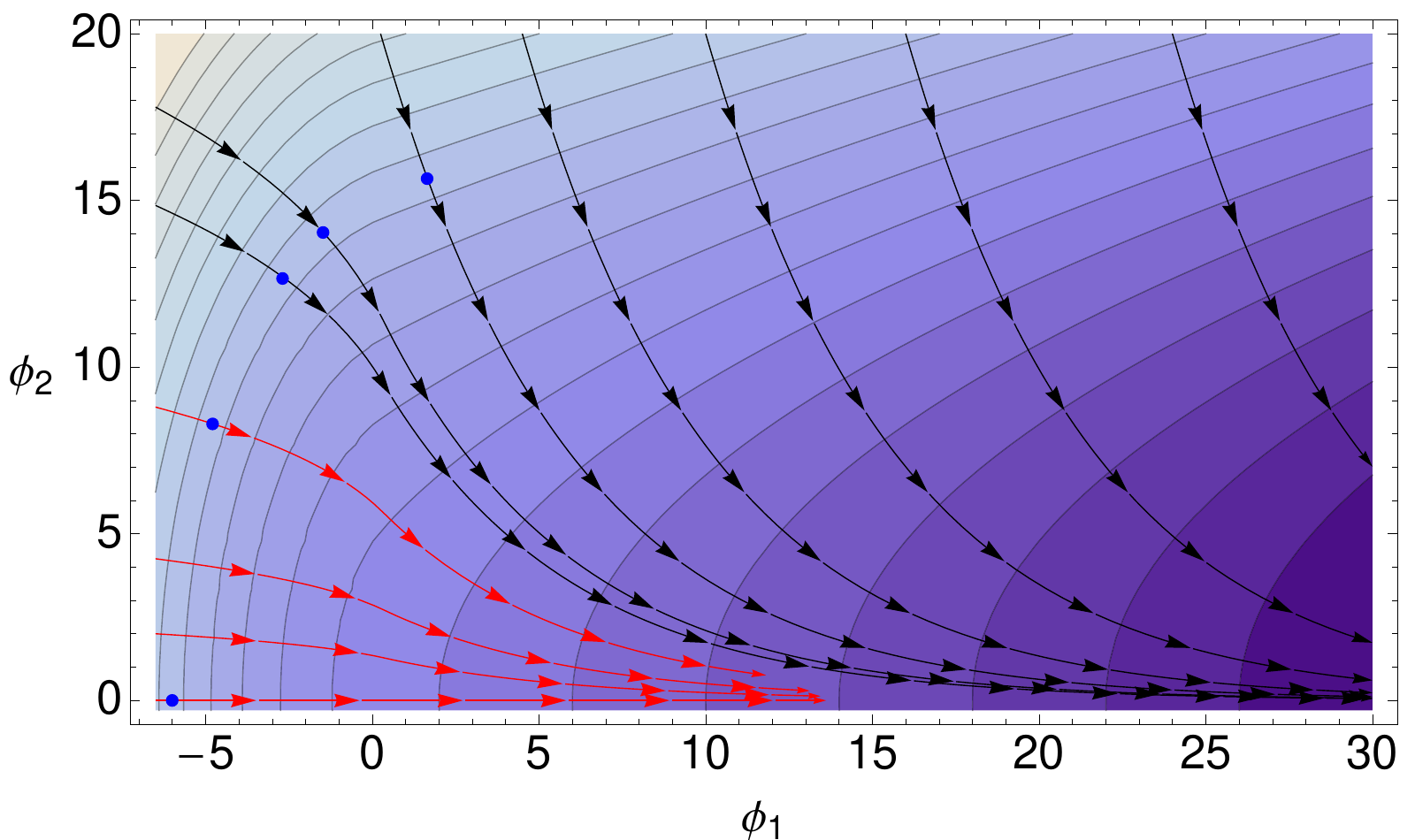}
  \hfill
  \includegraphics[width=.45\textwidth]{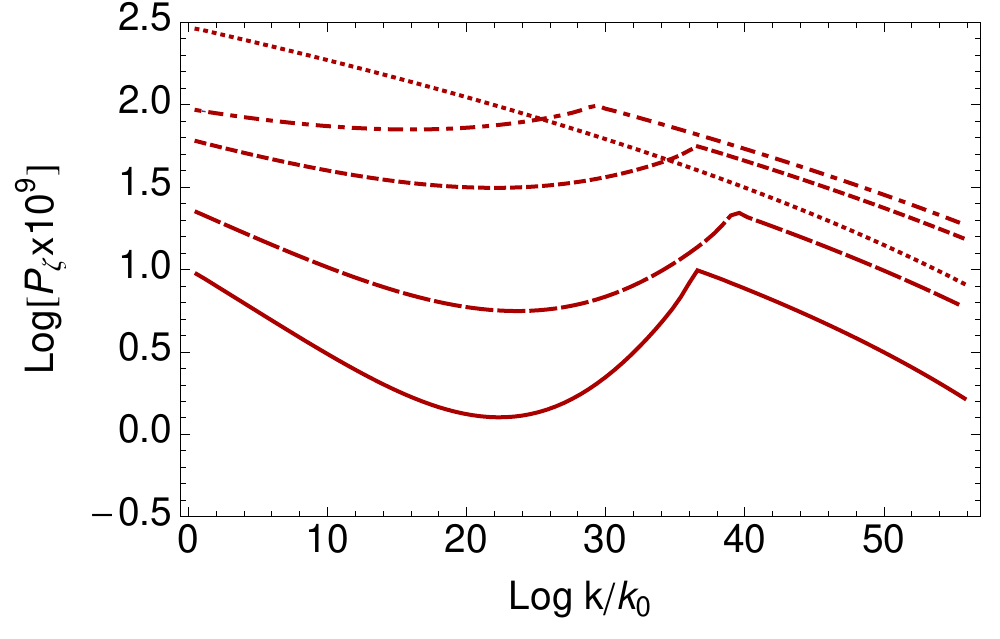} 
   \includegraphics[width=.48\textwidth]{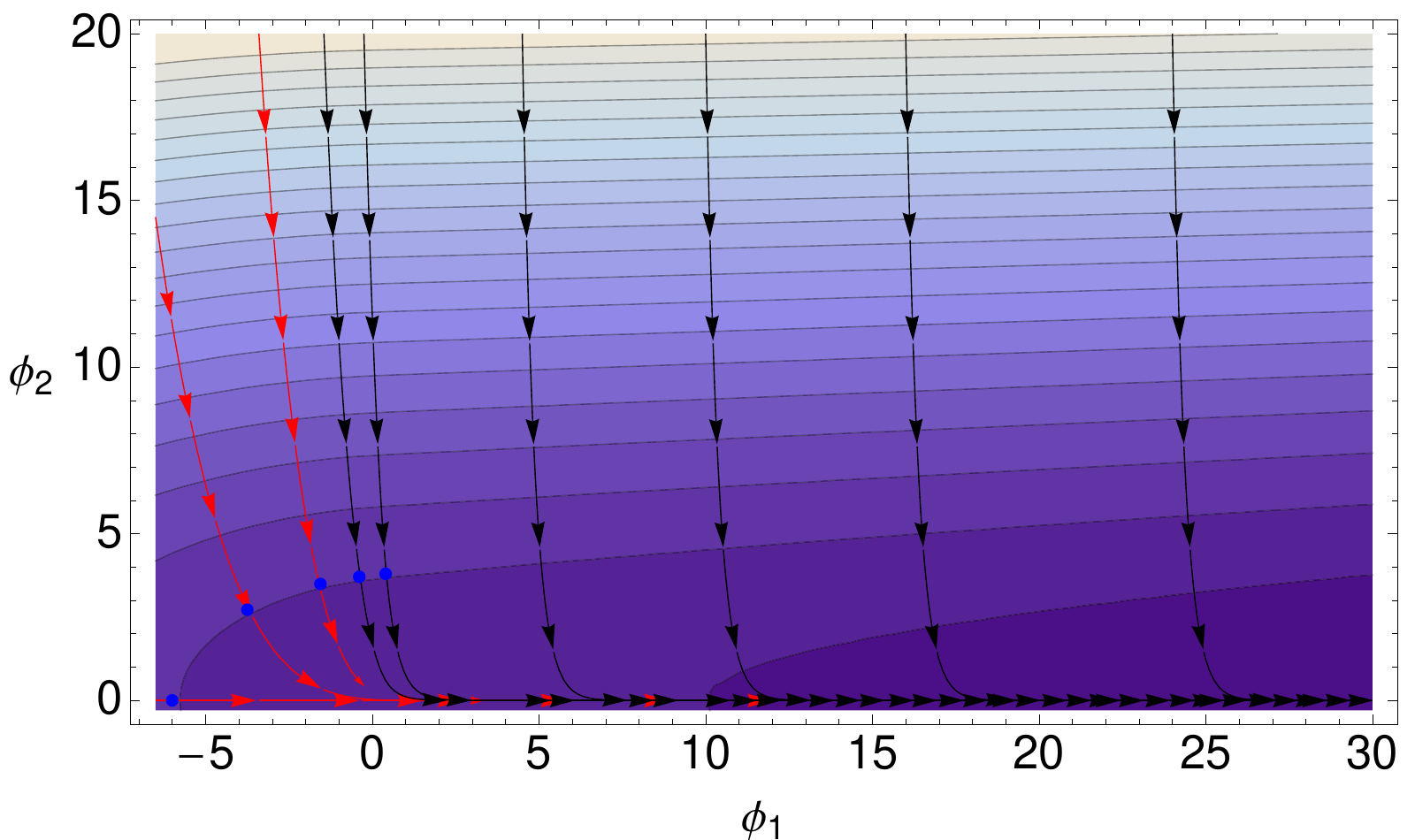}
	\hfill
   \includegraphics[width=.45\textwidth]{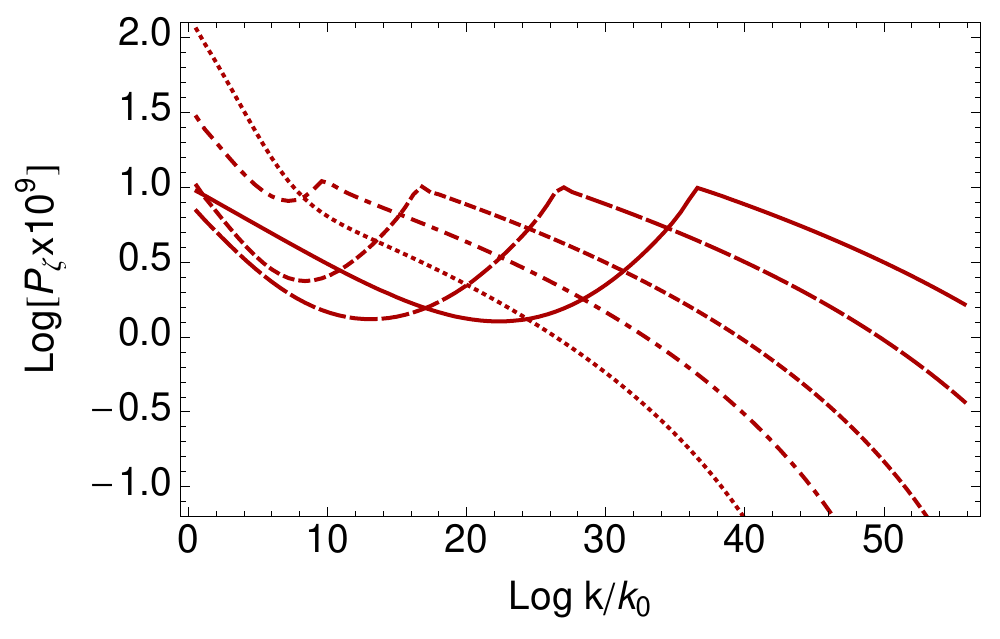}
  \caption{Scenario Two  $(m_{2}, m_{1} >0)$.  The  parameters of the model are set to $\epsilon_{\rm S}=0.005$, 
$\Lambda^4 = 3.5 \times 10^{-9}$ and $m_1^2 = 0.07$. The second  mass parameter  is  given, from top to bottom,  by $m_2^2= m_1^2, \, 0.25 \times m_1^2, \, 4 \times m_1^2$. LEFT: Inflationary trajectories  on the $\phi_1 - \phi_2$ plane.  The trajectories associated to a suppressed power spectrum are drawn in red. RIGHT:  Power spectra corresponding to the initial conditions represented  by  blue dots in the left   figures. From bottom to top the solid, 
long-dashed, short-dashed and  dotted lines correspond to increasing initial values for $\phi_2$ at $N=0$. To see more clearly the effect of changing the initial conditions, $k_0$ is fixed as in the left plot of Fig.~\ref{fig:Pofk_1d}.} 
\label{fig:m1Em2}
\end{figure}

\paragraph{Scenario Two $(m_{2}, m_{1} >0):$} In our second scenario we consider the effects of having a non-zero 
value for $m_{1}$. Taking $m_1$ to be the same as in the left-hand plot of Fig.~\ref{fig:Pofk_1d}, i.e.
appropriately chosen to manifest power suppression when the second field is placed at its minimum,  we conclude that the results depend heavily on the choice 
of initial conditions. In Fig.~\ref{fig:m1Em2} we show the inflationary trajectories and corresponding power 
spectra of the cases $m_1= m_2$, $m_1\gtrsim m_2$ and $m_1\lesssim m_2$ respectively. In these cases depending 
on the choice of initial conditions  the result interpolates between the full suppression  
of the single-field model and no suppression, when $\phi_{2}$ plays a significant role in the evolution, as occurs in the previous example.

The left-hand plots show the slow-roll inflationary trajectories for each case. The red lines correspond to 
inflationary trajectories where the corresponding power spectrum displays suppression. For convention,  we say that a power 
spectrum has suppression when the deficit of $\log P_\zeta$  four $e-$folds before transition  with respect to the transition itself~$N=N_c$
\be
\log P_\zeta|_{N_c}- \log P_\zeta|_{N_c-4},
\ee
is at least half of what it is obtained in the single-field case.
The plots illustrate that the angle $\theta_c$ increases with the magnitude of $m_2$, as expected from the formula \eqref{thetaC}.

\begin{figure}
  \centering
  \includegraphics[width=0.9\textwidth]{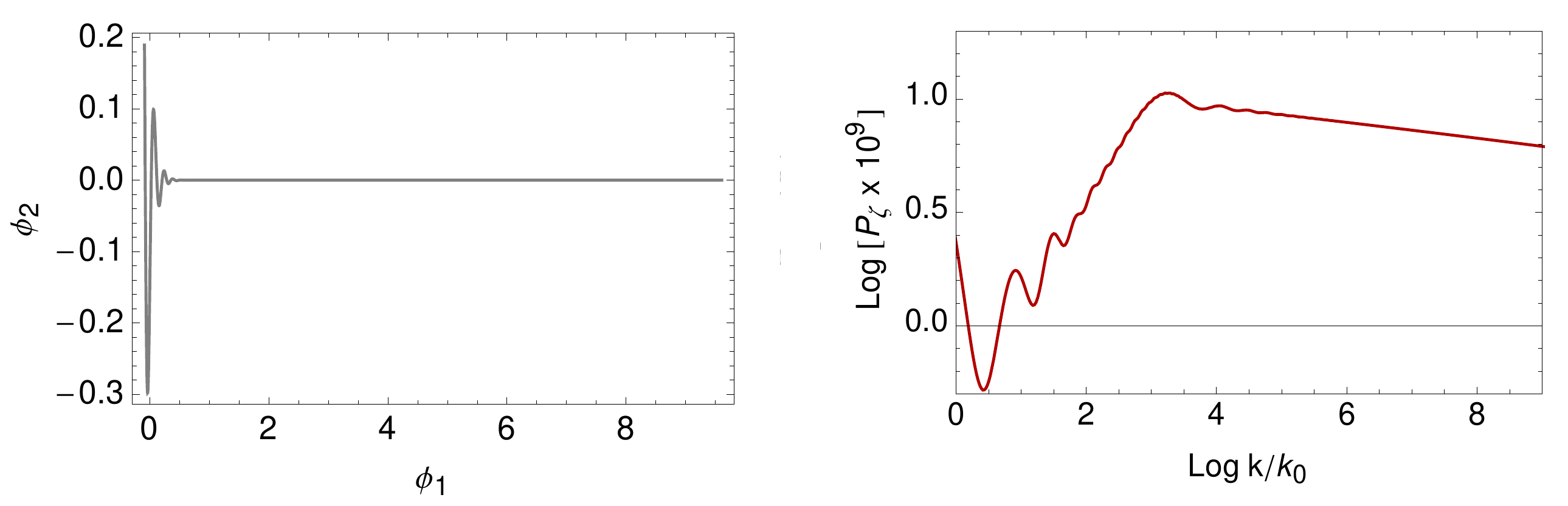}
 \caption{Scenario three with  $m_{1}^{2}=0.2$ and $m_{2}^{2}=40\times0.2$. LEFT: field trajectory; RIGHT: corresponding  $P_{\zeta}(k)$. The power suppression closely resembles the single-field 
example, except now there are also smaller scale oscillations resulting from the oscillations of the second field 
$\phi_{2}$. $k_0$, $\Lambda$ and $\epsilon_{\rm S}$ are defined as in Fig.~\ref{fig:massDep}.}\label{fig:moreexamples} 
\end{figure}

\paragraph{Scenario Three $(m_2 \gg m_1):$} For completeness, our third scenario is the same as the last one, 
except now $m_2 \gg m_1$. This case cannot be treated using the slow-roll approximations since displacement of the second field  
can give rise to significant oscillations. 
As expected from basic considerations of effective field theory, as the limit $m_2 \gg m_1$ is approached, the situation
 becomes simpler and the broad features match that of the single-field result. The difference is that now we also see 
oscillations on the largest scales as a result of excitation of the second field; see Fig.~\ref{fig:moreexamples}. This 
last scenario displays  essentially the heavy physics phenomenology that has been studied in, for example, Refs.~\cite{hep-th/0605045,1010.3693,1104.1323,1306.5680,1404.7522,1404.1536,1405.4257} and references therein.

The nature of the oscillations of a second field are different to that of the mixing of modes observed in the single-field 
scenario. It was shown in Refs.~\cite{1306.5680,1104.1323} that the features related to the decay of a heavy mode behave as
\begin{equation}
\left(\frac{\Delta P}{P_0}\right)_{\rm osc} \propto \frac{1}{k^3} \cos\left[ 2\frac{m_h}{H} \ln \left(\frac{k}{a_* m_h}\right)\right]
\end{equation}
where $m_h$ is the mass of the heavy field and $P_0$ is the unperturbed power spectrum. 
The mixing of modes as a result of a sudden change in the background evolution \cite{Martin:2011sn,astro-ph/0303636,Cicoli:2014bja} 
gives rise to a different oscillatory profile. Even though the amplitude of oscillations is sensitive to details of the period of 
rapid evolution, in the short wavelength limit it is possible to characterise their frequency and decay rate in a relatively 
model independent way.

Provided  the density perturbations asymptote to the Bunch-Davis vacuum for scales deep inside the horizon, the power spectrum has the following characteristic behaviour for  $k/a_cH \gg1$ (see Ref.~\cite{Martin:2011sn})
\begin{equation}
\left(\frac{\Delta P}{P_0}\right)_{\rm osc} \propto \frac{1}{k} \sin\left(\frac{2k}{a_c H}\right).
\end{equation}
where $a_c$ is the value of the scale factor at the transition $N=N_c$. With sufficiently high quality data one could hope to distinguish these two signatures.

\section{Example II: Power suppression from superhorizon evolution }

\label{SHsuppression}

So far all our examples have been simple extensions of a single-field model with an initially steeper phase 
of inflation. In the examples shown, multifield effects only act as to reduce the amount of power suppression 
since larger scales experience more superhorizon evolution than smaller scales. However, as described in 
\S~\ref{sec:suppresionMulti}, in principle it should be possible to generate power suppression on large 
scales purely as a consequence of smaller scales experiencing a greater level of superhorizon evolution. This 
is a very different mechanism to that of the previous sections and, to our knowledge, there are no examples 
of models of this kind in the literature. The purpose of this short section therefore is to provide such an example.

\begin{figure}
  \centering
  \includegraphics[width=0.4\textwidth]{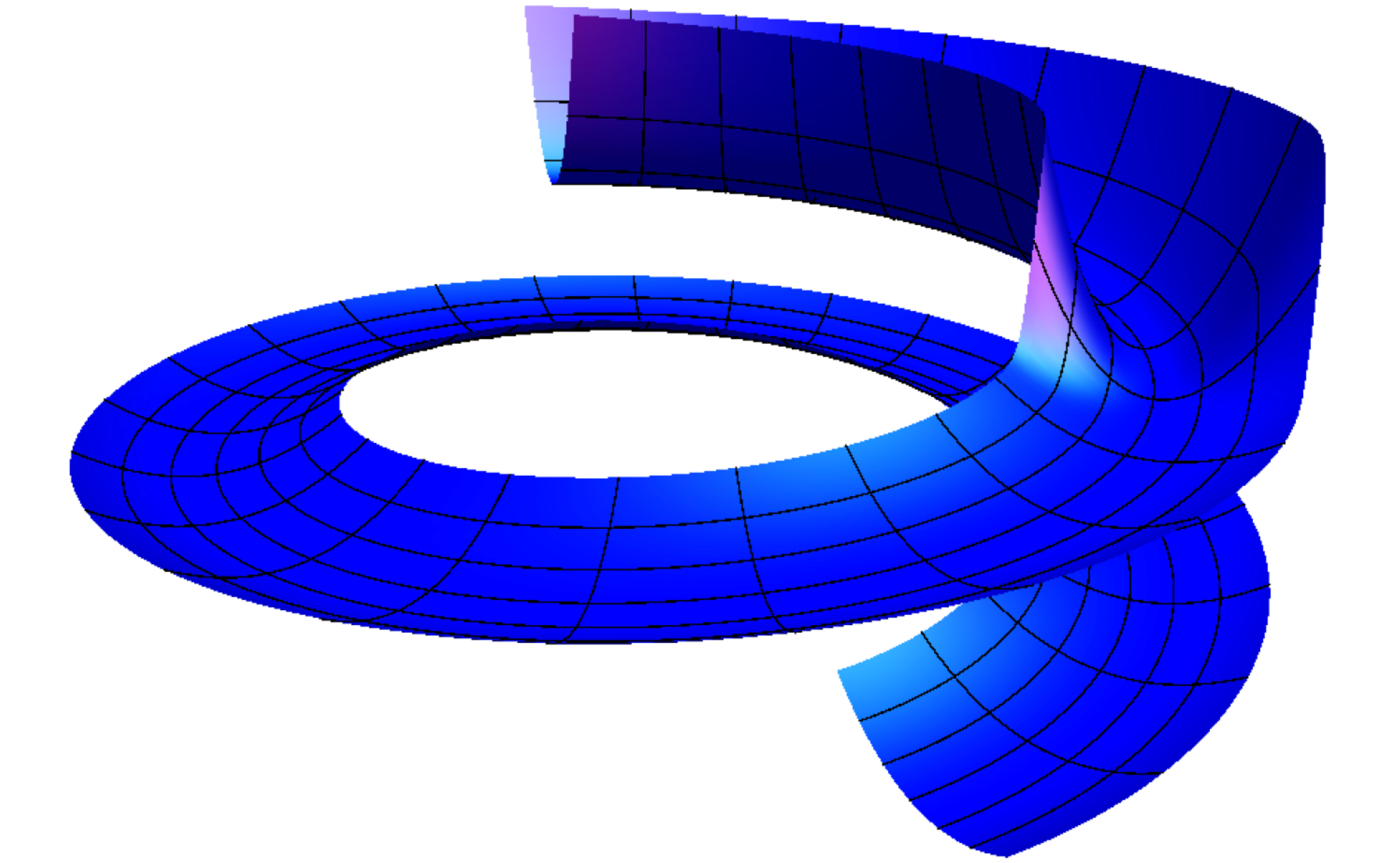}
    \caption{Scalar potential in flume inflation, defined by equations \eqref{flumeMetric} and \eqref{eq:flumePotential}. The inflaton rolls down the potential  mainly along the angular direction. The mass associated to the radial direction, $m^2$, changes abruptly  from a heavy phase where $m^2 \gg H^2$, and thus the model is effectively single field, to a light phase where $m^2 < H^2$, and therefore the effects  of superhorizon evolution of the perturbations become relevant.} \label{fig:FlumePotential}
\end{figure}

The model we propose is inspired by inflation in not single-valued potentials which are present in
descriptions of moduli spaces in string theory models of compactifications (see also \cite{Li:2014vpa, 1412.5093}). However, for the purposes of this 
paper, we will treat it purely as a toy model and make no claims of a particle physics embedding. This interesting 
possibility is left to future work. To be getting on with, we shall simply refer to this preliminary model
 as \emph{Flume Inflation}. The functional form of this model is very similar to the previous examples, 
except now we work in polar coordinates $\{r,\theta\}$. The field-space metric is thus
\begin{equation}
G_{ij} = \begin{pmatrix}
1 & 0 \\
0 & r^2
\end{pmatrix}
\label{flumeMetric},
\end{equation}
 but note that the field space  $\cM$ has zero curvature, and therefore geodesics are still straight lines in cartesian coordinates. 
 The potential has the form
\begin{equation}
V=\Lambda^{4}\left[1+\sqrt{2 \epsilon_{s}} \theta + \frac{1}{2}m(\theta)^{2}(r-r_{c})^{2}\right],
\label{eq:flumePotential}
\end{equation}
plotted in Fig.~\ref{fig:FlumePotential}, where the key step is to make the radial mass a function of $\theta$, such that it becomes lighter at some 
point $\theta_{c}$ while relevant scales are leaving the horizon. To achieve this in an intuitive way we use the error function
\begin{equation}
m(\theta)^{2}=m_{h}^{2}+\frac{1}{2}\left(1+\erf \left(\frac{\theta-\theta_{c}}{\Lambda_{h}}\right)\right)\left(m^{2}_{l}-m_{h}^{2}\right).
\end{equation}
where $\Lambda_{h}$ determines the length scale over which the mass changes from $m$ to $m_{l}$. Initially 
we set $m^{2}\approx m_{h}^{2}>H^{2}$ such that isocurvature modes exiting the horizon during this phase are 
rapidly decaying and hence the evolution, despite the curved trajectory, is essentially single field. After the 
transition, the mass rapidly approaches $m_{l}^{2}\ll H^{2}$ such that the slow-roll approximations are valid at 
late times and hence the model is amenable to the discussion in \S~\ref{sec:suppresionMulti}. 
During the transition from the heavy phase to the light phase the approximations of \S~\ref{sec:suppresionMulti} do
 not hold and hence the results shown are obtained numerically without making any slow-roll approximations. Nevertheless 
at a qualitative level we will see that the intuition obtained from \S~\ref{sec:suppresionMulti} serves us well.

\begin{figure}
  \centering
  \includegraphics[width=0.9\textwidth]{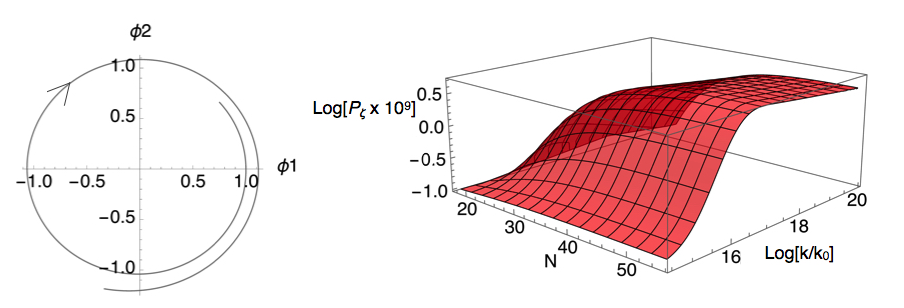}
    \caption{The lefthand plot shows the trajectory to spiral outwards as result of the decreasing radial mass. The righthand plot demonstrates that while the spectrum is close to flat at horizon crossing, superhorizon evolution causes the amplitude on smaller scales to grow significantly, resulting in a spectrum which is more commonly referred to as ``suppressed" on the largest scales.} \label{fig:FlumeExample}
\end{figure}

The dynamics of the inflaton is reminiscent of a person riding down a spiralling flume. Much like the person, the inflaton will 
sit close to the minimum in the radial direction, and will roll  mainly along the angular direction. 
This  deviation from the geodesics causes excitation of isocurvature modes and also transfers power from isocurvature 
perturbations to adiabatic perturbations, but this effect is only significant when  the radial mass becomes sufficiently light. 
We therefore expect $\Delta_N \neq 0$ for scales exiting the horizon during the later 
stages when $m^{2}\approx m_{l}^{2}$. Hence, by arranging for the radial mass to become lighter during the period when CMB 
scales are exiting the horizon, we ensure that smaller scales undergo a larger amount of superhorizon evolution than larger scales \footnote{If this behaviour continues until the end of inflation, then the adiabatic limit will not be reached and hence the model is in principle  sensitive to the details of reheating.  We will not address this possibility here.}. 

An example of this is shown in Fig.~\ref{fig:FlumeExample}. The lefthand plot shows that although the minimum is at a fixed 
radius, the trajectory spirals outwards as a result of the decreasing mass. The righthand plot shows that at horizon exit the 
power spectrum is close to flat. This is a direct consequence of $\epsilon \approx \text{constant}$. We then see that smaller scales 
undergo a substantial amount of superhorizon evolution while large scales do not. This is because the isocurvature on the
 largest scales underwent a period of exponential decay and hence, even when the radial mass becomes light, isocurvature 
modes never grow to the extent that isocurvature modes exiting the horizon during the lighter phase do. The result is that 
by the end of inflation the spectrum does indeed appear suppressed on large scales.

\section{Discussion}

Observing statistically significant power suppression on the largest observable scales would require some non-minimal 
version of inflation and therefore radically improve the possibility of extracting valuable information about fundamental 
physics from the CMB. A number of authors have pointed out that a steepening of the potential associated with a tunnelling 
event can result in suppression of the power spectrum on large scales. Our paper extends this analysis by discussing the 
implications of this in the context of multifield inflation. 

We find that the existence of a steepening along the inflationary direction is not a sufficient 
condition for power suppression, but rather that there is a relevant direction in field space that determines
whether or not the power spectrum presents this suppression. This direction depends on the full inflationary trajectory
in field space and is therefore not something determined locally from the potential. Furthermore, its orientation
does not have to be aligned with the velocity vector in field space so fast evolution does not necessarily lead to
power suppression even if it is present at horizon crossing. The reason for this behaviour is that the superhorizon
evolution of isocurvature modes can source the adiabatic mode, thereby restoring its level of power 
after horizon crossing.

For models where the potential can be expressed as sum-separable, the situation becomes sufficiently simple such that it is possible to give
 analytic expressions for the constraints on the potential at horizon crossing to ensure
large scale suppression of the power spectrum. This is further simplified in models
where the adiabatic limit is reached shortly after the transition from the steep to the flat part of the 
potential where most of the e-folds of inflation would take place. We present simple 2-field models 
where we can easily visualise the dynamics and show explicit examples where the suppression at horizon
crossing completely disappears after superhorizon evolution is taken into account. We also study
the resultant power spectrum as a function of the parameters of the potential and initial
conditions. This is important in the landscape picture where the initial conditions would be 
determined by an instanton that interpolates between the previous vacuum and the one described
by this potential.

We also describe a novel approach to power suppression where the superhorizon evolution of
perturbations plays a key role. One can
envision a model where the relative power suppression is created by an enhancement of the
level of the perturbations after the first few e-folds of inflation, in other words, one does not
suppress the power at large scales but enhances it at small scales. We present a toy model
that achieves this as a result of transfer of power from isocurvature to adiabatic perturbations over a
prolonged period. This example may have interesting
applications in inflationary models arising from compactification scenarios. 

Our results have interesting implications for inflation in the landscape. If our observable 
patch emerged after tunnelling from a metastable vacua, then the combined conditions 
on the potential of the active fields and their initial conditions in order to achieve a suppressed large scale power spectrum imposes strong constraints on where 
in the landscape this event might have happened. In summary, observation of power suppression on large scales would enable one to establish 
a connection between the initial conditions for the onset of inflation in the landscape and the dynamics 
of active fields. We have made first steps in making this relationship precise.

\section{Acknowledgements}
The authors wish to thank Ana Achucarro, Pablo Ortiz, Yvette Welling and Jon Urrestilla for helpful discussions.
 J.J.B.-P. and JF are supported by IKERBASQUE, the Basque Foundation for Science. J.J.B.-P. is also supported 
in part by the Spanish Ministry of Science grant (FPA2012-34456) and Consolider EPI CSD2010-00064. KS acknowledges financial support from the University of the Basque Country UPV/EHU via the programme ÒAyudas de Especializacion al Personal
Investigador (2012)Ó, from the Basque Government (IT-559-10), the Spanish Ministry
(FPA2009-10612) and the Spanish Consolider-Ingenio 2010 Programme CPAN (CSD2007-
00042) and Consolider EPI CSD2010-00064. M. D.  acknowledges support from the European Research Council under the European Unions Seventh Framework Programme (FP/20072013)/ERC Grant Agreement No. [308082]. We would particularly like to thank Baobab bar -- teteria in Bilbao, where a significant portion of this work was done.

\begin{appendix}

\section{Evolution of the scalar perturbations in the kinematical basis}

In this appendix we will derive the equation for the scalar perturbations on superhorizon scales in the kinematic basis, where the evolution of curvature and isocurvature perturbations is explicit. We follow the notation of Refs. \cite{1005.4056, 1011.6675, 1111.0927}. In these works, the curvature perturbation is defined in the comoving gauge and therefore denoted 
$\mathcal{R}$. Curvature perturbations defined in the constant density gauge --- $\zeta$ --- and in the comoving gauge --- $\mathcal{R}$ --- are known to be equal at second order on superhorizon scales up to $\mathcal{O}(k/aH)^{2}$ corrections \cite{0809.4944,astro-ph/0411463}. For consistency with the standard usage of the separate universe assumption, in this paper, we will always refer to the curvature perturbation as $\zeta$.

For simplicity we will just discuss the case where the field space metric is flat everywhere, and therefore will not make any distinction between upper and lower indices. Moreover, we restrict ourselves to the slow-roll regime, where the background satisfies the equation
\be
\boldsymbol{\phi'} = - \boldsymbol{\nabla} \log V.
\label{backEqApp}
\ee
In particular we will show that curvature perturbations are conserved on superhorizon scales  in the absence of entropy modes, and that  curvature perturbations are not transferred into entropy modes. This two facts were used in \S~\ref{background} to explain the simple structure of the transfer matrix $\boldsymbol{T}$ in \eqref{transfer matrix multi}. 

\subsection{Equations for the scalar perturbations}

As we discussed in \S~\ref{background}, scalar perturbations $ \boldsymbol{\delta \phi}$ on superhorizon scales satisfy
\be
\frac{D \boldsymbol{\delta \phi}}{dN} = - \boldsymbol{M}\cdot \boldsymbol{\delta \phi} \qquad \text{where} \qquad \boldsymbol{M} = \boldsymbol{\nabla}\boldsymbol{\nabla} \log V. 
\label{pertEqApp}
\ee
The evolution of scalar perturbations becomes particularly simple when the perturbations are decomposed in the kinematic basis, $B_k\equiv\{\mathbf{e}_{\parallel}, \mathbf{e}^\alpha_\perp\}$, which is adapted to the background evolution.  The basis vectors $\mathbf{e}_{\parallel}$ and $\mathbf{e}_{\perp}^{(2)}$ (we use $(2)$ to refer to the element $\alpha=2$, not to be confused by an exponent 2) are defined as 
\be
\mathbf{e_\parallel}\equiv \frac{ \boldsymbol{\phi'}}{v}, \qquad \frac{D \mathbf{e_\parallel}}{dN}\equiv Z_{21} \mathbf{e}^{(2)}_\perp \qquad \text{where} \qquad v \equiv \sqrt{2 \epsilon}.
\label{basisDefApp}
\ee
The vector $\mathbf{e}_\parallel$ has unit norm since $v = |\boldsymbol{\nabla}\log V| = |\boldsymbol{\phi}'|$, and  $Z_{21}$ is defined to be a real positive number.  The precise definitions of the rest of the basis vectors  $\mathbf{e}^\alpha_\perp$  with $\alpha>2$ will not be necessary for our derivation, and  can be found in Ref.~\cite{1111.0927}. For our purposes it suffices to  know that the full set $B_k$ forms an orthonormal basis, i.e.  $\mathbf{e}_a \cdot \mathbf{e}_b = \delta_{ab}$, where $ \mathbf{e}_a =\{ \mathbf{e}_\parallel , \mathbf{e}_\perp^\alpha \}$. As in the case of the vector $\mathbf{e}^{(2)}_\perp$, the evolution of the rest of the basis vectors along the inflationary trajectory is encoded in the elements of  matrix $\boldsymbol{Z}$ 
\be
\boldsymbol{Z}_{ab} = \mathbf{e}_a \cdot \frac{D \mathbf{e}_b}{dN}. 
\ee
It is easy to check that this matrix is antisymmetric from the  condition $\frac{d}{dN} (\mathbf{e}_a \cdot \mathbf{e}_ b )=0$, which follows from the orthonormality of this set of vectors. In the kinematic basis the perturbations 
$\boldsymbol{\delta \phi}$ can be decomposed as
\be
\boldsymbol{\delta \phi}= v \zeta \mathbf{e}_\parallel + \sum_{\alpha=2}^{\scN} \delta \phi_{\perp}^\alpha\,   \mathbf{e}_\perp^\alpha.
\label{decompApp}
\ee
As we mentioned above,  equations \eqref{pertEqApp} become particularly clear when the perturbations are decomposed in this way, and this is due to the  simple structure that the matrix $\boldsymbol{M}$ has on the basis $B_k$. To see this first note that from the background equation \eqref{backEqApp} we can derive the following two relations
\be
 \frac{D}{dN} = \boldsymbol{\phi}' \cdot \boldsymbol{\nabla} = - (\boldsymbol{\nabla} \log V) \cdot \boldsymbol{\nabla}  \quad \Longrightarrow \quad\frac{D \boldsymbol{\phi}'}{dN}   =  -\frac{D }{dN}( \boldsymbol{\nabla} \log V) = - \boldsymbol{M} \cdot \boldsymbol{\phi'}.
\ee
Then, using $\boldsymbol{\phi}' = v \mathbf{e}_\parallel$ and the definition for the basis vector $\mathbf{e}^{(2)}_{\perp}$ \eqref{basisDefApp}, we can rewrite the second equation as
\be
 \frac{D\boldsymbol{\phi}'}{dN} =  \frac{d v }{dN} \mathbf{e}_{\parallel} + v Z_{21} \mathbf{e}_{\perp}^{(2)} = - v \boldsymbol{M} \cdot \mathbf{e}_{\parallel} .
\ee
Projecting this relation along the vectors  $\mathbf{e}_a=\{\mathbf{e}_{\parallel}, \mathbf{e}^\alpha_\perp\}$ we find that the elements of the matrix $\boldsymbol{M}$ in this basis  satisfy
\be
\mathbf{e}_{\parallel} \cdot \boldsymbol{M} \cdot \mathbf{e}_{\parallel}= -\frac{1}{v} \frac{D v }{dN}, \qquad 
\mathbf{e}_{\perp}^{(2)} \cdot \boldsymbol{M} \cdot \mathbf{e}_{\parallel}= -Z_{21}, \qquad  
\mathbf{e}_{\perp}^\beta \cdot \boldsymbol{M} \cdot \mathbf{e}_{\parallel} = 0 \quad \text{for all} \quad \beta>2.
\ee
Finally, when the expression for the perturbations in the kinematic basis \eqref{decompApp} is substituted into  equation \eqref{pertEqApp} we find 
\be
\frac{d v}{dN}  \zeta \mathbf{e}_\parallel  + v \frac{d \zeta}{dN} \mathbf{e}_\parallel + v \zeta \frac{D \mathbf{e}_\parallel}{dN} +  \frac{d \delta \phi_{\perp}^\alpha}{dN} \mathbf{e}_\perp^\alpha +\delta \phi_{\perp}^\alpha  \frac{D \mathbf{e}_\perp^\alpha}{dN} =
 - v \zeta \boldsymbol{M} \cdot  \mathbf{e}_\parallel - \delta \phi_\perp^\alpha \, \boldsymbol{M} \cdot  \mathbf{e}_\perp^\alpha, 
\ee
and after projecting this expression along the  vectors $\mathbf{e}_\parallel$ and $\mathbf{e}_\perp^\alpha$, we find that the curvature perturbations $\zeta$  and isocurvature perturbations $\phi_\perp^\alpha$ satisfy respectively 
\bea
\label{A10}
\frac{d \zeta}{dN} &=& - 2 Z_{21} \frac{\delta \phi_\perp^{(2)}}{v},\nonumber \\
\frac{d \delta \phi_\perp^\alpha}{dN} &=& -[\boldsymbol{M}- \boldsymbol{Z}]_{\alpha\beta}\;  \delta \phi^\beta_\perp.
\eea
From this equations it is now clear that curvature perturbations are conserved in the absence of isocurvature pertubations, and that isocurvature is not sourced by curvature perturbations. It is straight forward to check that, when  the second equation is written in terms of the entropy perturbations 
$\cS^\alpha \equiv \delta \phi^\alpha/v$, it takes the form 
\be
\label{A11}
\frac{d \cS^\alpha}{dN} = -[\boldsymbol{M}- \boldsymbol{Z} - M_{\parallel \parallel}\,  \unity]_{\alpha\beta}\; \cS^\beta \qquad \text{where} \qquad  M_{\parallel \parallel}  \equiv \mathbf{e}_{\parallel} \cdot \boldsymbol{M} \cdot \mathbf{e}_{\parallel}.
\ee

\subsection{The transfer matrix}

The set of equations \eqref{A10} and \eqref{A11} can be written  in matrix form as follows 
\be
\label{matrixEqApp}
\frac{d}{dN}
\begin{pmatrix}
\zeta\\
\cS^\alpha
\end{pmatrix} =  
\begin{pmatrix}
0 & -2 Z_{21} \, \delta_{2\beta}\\
0& -[\boldsymbol{M}- \boldsymbol{Z}]_{\alpha\beta} - M_{\parallel \parallel}\,  \delta_{\alpha\beta}
\end{pmatrix}\cdot \begin{pmatrix}
\zeta\\
\cS^\beta
\end{pmatrix} 
\ee
The transfer matrix $\boldsymbol{T}(N_2,N_1)$ between two times labelled by the e-fold numbers $N_1 < N_2$ is defined as
\be
\begin{pmatrix}
\zeta\\
\cS^\alpha
\end{pmatrix}_{N_2}  = \boldsymbol{T}(N_2,N_1) \cdot \begin{pmatrix}
\zeta\\
\cS^\beta
\end{pmatrix}_{N_1} 
\ee
And therefore, the equation for the perturbations \eqref{matrixEqApp} implies that the transfer matrix between the times $N$, and $N+dN$  is given by
\be
\label{transferInfApp}
\boldsymbol{T}(N + dN,N) =  \begin{pmatrix}
1 & -2 Z_{21} \, \delta_{2\beta}\\
0&(1- M_{\parallel \parallel})\,  \delta_{\alpha\beta}  -[\boldsymbol{M}- \boldsymbol{Z}]_{\alpha\beta}
\end{pmatrix}
\ee
 Since,  by definition, transfer matrices satisfy the composition rule
 \be
  \label{compositionApp}
  \boldsymbol{T}(N_3,N_1)= \boldsymbol{T}(N_3,N_2) \cdot \boldsymbol{ T}(N_2,N_1),
    \ee
the full transfer matrix $ \boldsymbol{T}(N_2,N_1)$ can be obtained formally by multiplying a sequence of infinitesimal  transfer matrices of the form   \eqref{transferInfApp}. Moreover, as the composition preserves the upper triangular structure in \eqref{transferInfApp} and also that $ \boldsymbol{T}_{\zeta\zeta}=1$, it follows that  the structure of the full transfer mass matrix $\boldsymbol{T}(N_2,N_1)$,  has   to be of the form
\be
\boldsymbol{T} = \begin{pmatrix}
1 & \boldsymbol{T}_{\perp \zeta}\\
0 & \boldsymbol{T}_{\perp \perp}
\end{pmatrix}.
\ee

\section{Spectral index in multifield inflation}
For convenience we define the transfer matrix $ \boldsymbol{\tilde T}$  as
\be
 \boldsymbol{\tilde T} \equiv \frac{v}{v^*}  \boldsymbol{T} \qquad \Longrightarrow \qquad  \boldsymbol{\delta \phi}(N) = \boldsymbol{\tilde T}(N,N_*) \cdot \boldsymbol{\delta \phi_*},
\ee
which satisfies the same formal equation as $\boldsymbol{\delta \phi}$
\be
 \qquad \frac{D \boldsymbol{\tilde T} }{d N} (N,N_*)= - \boldsymbol{\tilde M} \cdot \boldsymbol{\tilde T},
\label{Tequation}
\ee
Transfer matrices $\boldsymbol{\tilde T}$ satisfy the composition rule \eqref{compositionApp}, which in particular implies that 
  \be
\boldsymbol{\tilde T}(N,N_*) \cdot \boldsymbol{\tilde T}(N_*,N) =  \mathbb{I}.  
  \ee
Taking a derivative with respect to $N_*$ of the previous expression we find
\be
\frac{D \boldsymbol{\tilde T}(N,N_*)}{d N_*} \cdot \boldsymbol{\tilde T}(N_*,N)  = -  \boldsymbol{\tilde T}(N,N_*) \cdot \frac{D \boldsymbol{\tilde T}(N_*,N)}{d N_*}.
\ee
Then, using the equation for $\boldsymbol{\tilde T}$ (\ref{Tequation}), and  projecting along the inflationary direction at the end of inflation, $ \mathbf{e_{\parallel}}(N)$  we arrive to 
\be
  \frac{D \boldsymbol{\tilde T} }{d N^*}(N,N_*) =   \boldsymbol{\tilde T} \cdot \boldsymbol{\tilde M}, \quad \Longrightarrow \quad \frac{D \boldsymbol{\tilde T_{\parallel}} }{d N^*} (N,N_*) =  \boldsymbol{\tilde T_{\parallel}} \cdot \boldsymbol{\tilde M},
\ee
where  we have  defined $\boldsymbol{\tilde T_{\parallel}} \equiv \mathbf{e_{\parallel}}^\dag(N) \cdot \boldsymbol{\tilde T}$. 
Finally, writing the power spectrum   as  
\be
\zeta =\frac{1}{v} \boldsymbol{\tilde T_{\parallel}} \cdot \boldsymbol{\delta \phi_*}\qquad\Longrightarrow \qquad  P_{\zeta} = \left(\frac{\, H_*}{2 \pi v} \right)^2  \boldsymbol{\tilde T_{\parallel}} \cdot \boldsymbol{\tilde T_{\parallel}},
\ee
and given that the first equation also implies  $\mathbf{e}_{N} = \boldsymbol{\tilde T_{\parallel}}/|\boldsymbol{\tilde T_{\parallel}}|$, we have 
\be
\frac{d \log P_{\zeta}}{ d  N^*} = -2 \epsilon_* +  \frac{2}{|\boldsymbol{\tilde T_{\parallel}}|}\frac{D \boldsymbol{\tilde T_{\parallel}}}{d N_*} \cdot \mathbf{e}_{N} =  -2 \epsilon_* + 2  \mathbf{e}_{N}^\dag\cdot \boldsymbol{\tilde M}\cdot \mathbf{e}_{N}.
\ee

\end{appendix}

\bibliography{References}

\end{document}